\documentstyle[12pt]{article}
\def\beq{\begin{equation}}
\def\eeq{\end{equation}}
\def\bea{\begin{eqnarray}}
\def\eea{\end{eqnarray}}
\def\bq{\begin{quote}}
\def\eq{\end{quote}}

\def\CQG{{\it Class. Quantum Grav.} }

\def\IJMP{{\it Int. J. Mod. Phys.} }

\def\NP{{\it Nucl. Phys.} }
\def\PL{{\it Phys. Lett.} }
\def\PR{{\it Phys. Rev.} }

\def\ZP{{\it Z. Phys.} }

\def\gappeq{\mathrel{\rlap {\raise.5ex\hbox{$>$}}
{\lower.5ex\hbox{$\sim$}}}}

\def\lappeq{\mathrel{\rlap{\raise.5ex\hbox{$<$}}
{\lower.5ex\hbox{$\sim$}}}}

\begin{document}
\pagestyle{empty}
\begin{flushright}
CERN-TH.7241/94 \\
DAMTP R-94/22
\end{flushright}

\begin{center}
{\bf DECOHERENCE OF FRIEDMANN-ROBERTSON-WALKER GEOMETRIES} \\
{\bf IN THE PRESENCE OF MASSIVE VECTOR FIELDS WITH U(1) OR SO(3) GLOBAL
SYMMETRIES} \\
\vspace*{0.5cm}
{\bf O. Bertolami}$^{*)}$ \\

Theory Division, CERN \\
CH - 1211 Geneva 23, Switzerland \\
\vspace*{0.2cm}
and \\
\vspace*{0.2cm}
{\bf P.V. Moniz}$^{+)}$ \\
University of Cambridge, DAMTP \\
Silver Street, Cambridge, CB3 9EW, UK \\
\vspace*{0.5cm}
{\bf ABSTRACT} \\ \end{center}
\vspace*{5mm}
\indent
Retrieval of classical behaviour in quantum cosmology is usually
discussed in
the framework of {\em midi}superspace models in the presence of scalar
fields
and the inhomogeneous modes corresponding either to gravitational or
scalar fields. In this work, we propose an alternative model to study
the
decoherence of homogeneous and isotropic geometries where the scalar
field is
replaced by a massive vector field with a global internal symmetry.
We study here the cases with $U(1)$ and $SO(3)$ global
internal symmetries. The presence of
a mass term breaks the conformal invariance and allows for
the longitudinal modes of the spin-1 field to be present in the
Wheeler-DeWitt
equation.
In the case of the U(1) global internal symmetry, we have only one
single
 ``classical'' degree of
freedom while in the case of the SO(3) global symmetry, we are led to
consider a
 simple two-dimensional
minisuperspace model. These minisuperspaces are shown to be
equivalent to a set of coupled harmonic oscillators
where the kinetic term of the longitudinal modes has a coefficient
proportional
to the inverse of the scale factor. The
conditions for a suitable decoherence process and correlations
between coordinate and momenta are established. The validity of the
semi-classical Einstein equations  when
massive vector fields (Abelian and non-Abelian) are present is also
discussed.

\vspace*{0.2cm}
\noindent
PACS: 04.60.+n,98.80.-Cq,-Dr

\vspace*{0.3cm}
\noindent  \rule[.1in]{16.5cm}{.002in}

\noindent
$^{*)}$ On leave from Departamento de F\'{\i}sica, Instituto Superior
T\'ecnico,
Av. Rovisco Pais, 1096 Lisboa, Portugal. Present address: INFN - Sezione
Torino, Via Pietro Giuria 1, 10125 Turin, Italy; e-mail:
Orfeu@vxcern.cern.ch and Bertolami@toux40.to.infn.it

\noindent
$^{+)}$ Work supported by a Human Capital and Mobility Fellowship from
the
 European
Union (Contract ERBCHBICT930781); e-mail: prlvm10@amtp.cam.ac.uk
\vspace*{0.2cm}

\begin{flushleft} CERN-TH.7241/94 \\
DAMTP R-94/22\\
October 1994
\end{flushleft}
\vfill\eject

\setcounter{page}{1}
\pagestyle{plain}

\section{Introduction}

\indent

The emergence of classical properties from the quantum mechanics
formalism
is still largely an open problem. Some progress has, however, been
achieved
through the so-called decoherence approach.  On
fairly general grounds, the decoherence process takes place  as one
considers the
system under study to be part of a more complex world that interacts
with
other subsystems, usually referred to as the ``environment". The latter
usually consists in
the set of unobserved or irrelevant degrees of freedom which
 are traced out, implying that, at
least in an operational
way,  the wave-function evolves non-linearly and is lead to its
collapse \cite{aaa}. In this way, quantum
interference effects among states of the system are
suppressed by the interaction with
the environment. This suppression comes about as one integrates out the
irrelevant degrees of freedom. This coarse-graining procedure leads to
an
effective action of the original model and clearly generalizes the usual
field fluctuation splitting that is adopted when carrying out a quantum
loop expansion in the standard effective action calculation. From the
operational point of view, decoherence in the context of quantum
cosmology,
implies that the
Universe is essentially an open system as
observers have necessarily  to disregard large
classes of variables in any relevant observation.
In addition to the notion of decoherence, which strictly means that
there is no
interference among different quantum states, a further condition that a
system must satisfy to be regarded as classical is, of course, that it
is driven
by classical laws, implying that a sharp correlation
between configuration space coordinates and conjugate momenta should
exist
in the wave function.

These ideas have been developed in some depth in the
context of quantum cosmological models
\cite{bb}--\cite{kk},
where considering the decoherence process is mandatory as the
observable Universe behaves clearly in a classical fashion and one
expects that
this classical features arise from a more fundamental quantum
description.
Furthermore, although minisuperspace models can be justified on symmetry
grounds, the truncation which turns the full quantum gravity problem
with its
infinite degrees of freedom into a problem with a finite number of
degrees of freedom actually
violates the uncertainty principle, as the amplitudes and momenta of
inhomogeneous modes are set to zero and the non-linear interactions of
those modes among themselves and with the minisuperspace degrees of
freedom are
disregarded. Moreover, the validity of the minisuperspace approximation
(cf.  \cite{hh} and references therein) is ensured only when the
back-reaction of the inhomogeneous modes on the minisuperspace variables
is shown to be small.

Most of the literature concerning the emergence of classical behaviour
in the
context of quantum cosmology   considers scalar fields
\cite{bb}--\cite{jj}, where the  environment corresponds
to the  inhomogeneous modes either of the
gravitational or of the scalar fields \cite{dd}--\cite{kk}. This is a
shortcoming of most of the decohering models discussed so far as prior
to the
inflationary epoch the Universe was dominated by radiation, i.e.
Yang-Mills
fields, and, after spontaneous symmetry breaking phenomena, by massive
vector
fields. Furthermore, a relevant issue concerning the decoherence
approach is
to stablish to what extent some of its features are specifically related
with scalar fields and to achieve this goal one has to consider other
fields to
 play the
role of environment.

In this paper we propose a model to study the decoherence of
homogeneous and isotropic geometries, where the  scalar field is
replaced by a massive vector field with a global internal symmetry.
Our aim is to assess if such models are on the same footing with the
ones
where the quantum to classical transition is achieved via tracing out
the higher
modes of scalar fields, or in other words, if massive vector fields are
equally effective in playing the role of
an environment with respect to the observed ``classical'' degrees of
fredoom.
Notice that the presence of
a mass term is a crucial feature as it breaks the conformal symmetry of
the
spin-1 field
action which leads to a Wheeler-DeWitt equation where the gravitational
and
matter degrees of freedom decouple \cite{lll}, similarly to the case of
a free
massless conformally invariant scalar field \cite{mm}. Hence, the
presence of a
mass term allows for interaction between gravitational and matter
degrees of freedom providing a scale at which the decoherence process
can take
place. Moreover, the breaking of conformal invariance makes it possible
for the
longitudinal modes of the spin-1 field to be present
in the Wheeler-DeWitt equation.
As we shall see, the resulting system is equivalent to a set of coupled
harmonic oscillators
where the kinetic energy terms of the longitudinal modes have a
coefficient proportional to the inverse of the scale factor.

An important point related to the discussion of quantum cosmology with
vector
fields
concerns the compatibility of the simple homogeneous and isotropic
Friedmann-Robertson-Walker (FRW) geometries we shall consider and the
matter content of the Universe.
As far as the homogeneous modes of the
vector fields
are concerned, we shall use here the Ansatz formulated in Refs.
\cite{lll},
\cite{nn}--\cite{ss1}. More specifically, since the physical
observables have to be $SO(4)$-invariant, the isometry group of closed
FRW cosmologies, the
fields with internal degrees of freedom may be allowed to
transform under $SO(4)$ if these
transformations can be compensated by an internal symmetry
transformation.
Fortunately, there is a large class of fields satisfying the above
conditions, namely the so-called {\it $SO(4)$--symmetric
fields}, i.e. fields which are invariant up to an internal symmetry
transformation. However, such construction can only be made
possible in the presence of non-Abelian
symmetries and hence, for the the Abelian
case, we shall require the homogeneous modes of
the spatial components of the vector field to vanish.
For a ${\bf R}\times S^3$ topology, such minisuperspace
constructions with homogeneous and isotropic metric and gauge fields
with
$SO(N)$ and $SU(M)$ ($N \geq 2$, $M \geq 3$)
gauge groups have been recently carried out. For Euclidean FRW
geometries  Einstein-Yang-Mills wormhole solutions have been found in
Ref. \cite{nn}.
Classical Einstein-Yang-Mills solutions for closed geometries and
$SO(N)$
gauge groups were obtained in Ref. \cite{oo}. The ground-state
wave-function
of the Einstein-Yang-Mills system with $SO(3)$ gauge group was found in
Ref.
\cite{lll}. Massive vector fields with $SO(3)$ global symmetry in flat
FRW
universes were studied in Ref. \cite{pp}. Finally, in a
${\bf R}^4$ topology, homogeneous
and isotropic metric and Yang-Mills field configurations were considered
in
Ref. \cite{qq} and the inclusion of the dilaton in the context of string
theories was carried out in Ref. \cite{rr}.

In this work, we shall study the decoherence process for homogeneous and
isotropic metrics in the presence of massive vector fields with $U(1)$
and
$SO(3)$ global symmetries in a ${\bf R}\times S^3$ topology.
For simplicity, we begin by considering the case with $U(1)$ global
symmetry. The role of the environment will be played by the
inhomogeneous modes of the $U(1)$ field and the corresponding
minisuperspace
will be actually one-dimensional as it has only a
single ``classical'' degree of freedom as physical observable,
the scale factor.
Although this choice of the matter
content may seem rather restrictive,
we shall see that some of  our results may be extended to
the richer and more interesting system
with $SO(3)$ global symmetry, which we analyse afterwards.
In particular, we shall establish,
 for both cases, the conditions
required for achieving correlation between coordinates and momenta and a
satisfactory decoherence process. The validity of the semi-classical
Einstein equations when
massive vector fields are present will be also discussed. The
minisuperspace for the non-Abelian model
hereby studied is two-dimensional due to the specific
Ansatz for the homogeneous modes of the spin-1 field (see Refs.
 \cite{nn}--\cite{ss1}).

Multi-dimensional minisuperspace models have of course,
a much richer structure and are therefore
far more interesting to consider in what regards the retrieval of
classical behaviour \cite{ggg,ggga,llo,cm,cm1}. In models containing
one single ``classical'' degree of freedom, the Hamilton-Jacobi
equation has only two solutions, generating the same trajectory
in opposite directions. The semi-classical wave-function
has two WKB components, each of
which may be called a WKB branch, of the form
\begin{equation}
\Psi[\mbox{$\cal{O}$},\mbox{$\cal{E}$}] = \sum_{(n)}
C_{(n)}[\mbox{$\cal{O}$}] e^{iM_P^2 S_{(n)}[\mbox{$\cal{O}$}]}
\psi_{(n)}[\mbox{$\cal{O}$}, \mbox{$\cal{E}$}].
\label{wkbinf}
\end{equation}
Here the subindex $(n)$ labels the WKB branches
(taking only two values, say $\pm 1$, uniquely identifying
the two possible solutions of the Hamilton-Jacobi equation)
and $\cal{O},\cal{E}$ denote, respectively, the ``classical''
physical observables and the extra degrees of freedom
corresponding to the environment. To achieve a proper decoherence, one
needs
not only that the
reduced density matrix turns out to be diagonal, but also that
the different WKB components in (\ref{wkbinf}) have negligible
interference among the diagonal terms. It is important to
stress that the analysis of correlations should be done
within each classical WKB branch (i.e. a diagonal term, $n=n^\prime$).
The interference
effects between the two possibilities of moving along the
one-dimensional
classical trajectory (corresponding to the expanding and collapsing
wave function components, respectively) were
shown to be effectivelly suppressed \cite{dd}, which is interpreted as
particle
creation \cite{cm}.

For a system with N degrees of freedom,
the Hamilton-Jacobi equation is expected to have an N-1 parameter
family of solutions, each one generating a N-1 parameter family of
classical
trajectories in the minisuperspace. In the multi-dimensional case, a
general
solution of the Wheeler-DeWitt equation may contain an {\em infinite}
superposition of semi-classical solutions of the form (\ref{wkbinf})
with
the subindex $(n)$ now corresponding to a set of parameters
that
uniquely identify a specific Hamilton-Jacobi solution.
However, each WKB branch must actually   be interpreted as
describing a whole family of classical trajectories, i.e. a set of
different universes  and
not a single one as for the N = 1 case.
Furthermore, the
N = 2 and N $>$ 2 cases are rather different as far as the
diagonalization of the reduced density matrix is concerned. These
differences
seem to have gone unnoticed until they were discussed in Ref.
\cite{ggga}. An
 example of N = 2 model
was studied in Ref. \cite{llo}, where the two ``classical'' degrees of
freedom
correspond to the scale factor and homogeneous mode of a
minimally massless scalar field and the environment was identified
with the inhomogeneous perturbations of {\em another}
minimally massless scalar field.
 The multi-dimensional cases
also allow one to better address the relation between  the
reduced density matrix formalism and the Feynman-Vernon influence
functional
\cite{fv}, and to the Schwinger-Keldish or
closed time path effective action \cite{sk}, as pointed out in
Ref. \cite{kiefer}.

Some preliminary work performed in Ref. \cite{ss}, with homogeneous and
 isotropic
cosmologies in the presence of
massive vector fields, has shown that
the necessary ingredients for  the process of decoherence to take place
are
present, although several aspects of the full
hyperspherical harmonics expansion of the fields remained
 to be fully assessed. The
Wheeler-DeWitt equation obtained in Ref. \cite{ss} is fairly similar to
the one
corresponding to a FRW minisuperspace model with a massive scalar field
with a
quartic self-coupling, $\lambda\phi^4$, conformally coupled with gravity
 \cite{hh}. In addition, we shall compare
the results of this paper with the ones obtained
previously in Refs. \cite{ggg,ggga}. Actually, there are many
similarities and this will allows
us to use some of the framework described in those
references\footnote{The
 authors
are grateful to B.L. Hu and J.P. Paz for pointing this out.}.
We would like to mention that, in considering perturbations in the
quantum
Einstein-Yang-Mills model of Ref. \cite{lll}, the authors of Ref.
\cite{uu}
consider also, as we do,
the harmonic expansion of the fields involved on $SO(4)$ and
$SO(4)/SO(3)$
(see above).

This paper is organized as follows. In Section 2
we present our model and, through general considerations, introduce an
Ansatz
for the vector fields with  global symmetry
as well as the expansion in $S^3$ hyperharmonics which will give rise
to an effective action.
We then proceed to a {\em midi}superspace description of such a model,
which will allow us to study both the $U(1)$ and $SO(3)$ global symmetry
cases.
In Section 3, we discuss the decoherence process and correlations
 within the $U(1)$  model and, in Section 4, we address the perturbed
 minisuperspace model with
massive vector fields with $SO(3)$ global symmetry. Our conclusions are
presented in Section 5.

\section{FRW Minisuperspace with Massive Vector Fields}

\indent

The action of our model consists of a Proca field coupled with gravity:
\beq
S = \int d^4x \sqrt{-g}~~\left[ {1\over 2k^2} (R - 2\Lambda) + {1\over
8e^2}
{\rm Tr}(\mbox{\boldmath $F$}^{(a)}_{\mu\nu}\mbox{\boldmath
$F$}^{(a)\mu\nu}) +
 {1\over 2}~m^2
{\rm Tr}(\mbox{\boldmath $A$}^{(a)}_{\mu} \mbox{\boldmath
 $A$}^{(a)\mu})\right]~,
\label{1}
\eeq
where $k^2 = 8\pi M^{-2}_P$, $M_P$ being the Planck mass, $e$ is a gauge
coupling constant  and $m$ the mass of the Proca field. To action
(\ref{1}) one
 adds the boundary action
\beq
S_B = -{1\over k^2} \int_{\partial M} d^3x \sqrt{h} K~,
\label{4}
\eeq
with $h_{ij} (i,j = 1,2,3)$ being the induced metric on the
three-dimensional
boundary $\partial M$ of $M$, $h = \det (h_{ij})$ and $K = K^{\mu}_\mu$
is the
trace of the second fundamental form on $\partial M$.

In quantum cosmology one is concerned with spatially compact topologies
 and we will consider here
the FRW Ansatz for  the  ${\bf R}\times S^3$  geometry
\beq
ds^2 = \sigma^2 a^2(\eta) \left[ -N(\eta)^2 d\eta^2 + \sum^3_{i=1}
\omega^i\omega^i\right]~,
\label{2}
\eeq
 where\footnote{Usually one makes the choice
$\sigma^2 = 2/3\pi M^2_P$, but  we shall keep the powers of $M^2_P$
explicitely in our effective action in order to compare our results with
the
ones of Refs. \cite{ggg,ggga,lll}.}
 $\sigma^2 = 2/3\pi$, $\eta$  is the conformal time,
$N(\eta)$ and $a(\eta)$ being the lapse function
and the scale factor, respectively and $\omega^i$ are left-invariant
one-forms in $SU(2)\simeq$ $S^3$ which satisfy
\beq
d\omega^k = - \epsilon_{kij} \omega^i \wedge\omega^j~.
\label{3}
\eeq

Aiming to obtain solutions of the Wheeler-DeWitt equation satisfied by
the wave
function $\Psi [h_{ij},\mbox{\boldmath $A$}^{(a)}_{\mu}]$:
\beq
\left[ G_{ijk\ell}~{\delta^2\over\delta h_{ij}\delta h_{k\ell}} + M^4_P
\sqrt{h}~(^{(3)}R - 2 \Lambda)   + {M^2_P\over 2} \sqrt{h}~T^{ii} \left[
\mbox{\boldmath $A$}_{\mu}^{(a)},-i{\delta\over\delta \mbox{\boldmath
 $A$}_{\mu}^{(a)}}\right]
\right] ~\Psi[h_{ij},\mbox{\boldmath $A$}_{\mu}^{(a)}]
 = 0~,
\label{5}
\eeq
where the superspace metric is given by
\beq
G_{ijk\ell} = {\sqrt{h}\over 2}~~(h_{ik} h_{j\ell} + h_{i\ell}h_{jk} -
h _{ij} h_{k\ell})~,
\label{6}
\eeq we shall expand the metric as
\beq
h_{ij} = \sigma^2a^2 ~~(\Omega_{ij} +\epsilon_{ij})~,
\label{7}
\eeq
with $\Omega_{ij}$ being the metric on the unit $S^3$ and
$\epsilon_{ij}$ a
perturbation that can be expanded  in scalar harmonics
${\cal D}^{J{\phantom J}M}_{{\phantom J}N}(g)$,
which are the usual $(2J+1)$-dimensional $SU(2)$
matrix representation and spin-2 hyperspherical harmonics
$Y^{2~LJ}_{m~MN}(g)$
on $S^3$ \cite{kk} as:
\beq
\epsilon_{ij} = \Omega_{ij} \sum_{J} {\sqrt{\overline{n}\over 3\pi^2}}
a_{J{\phantom J}M}^{{\phantom J}N}(\eta)
{\cal D}^{J{\phantom J}M}_{{\phantom J}N}(g) +
 \sigma^m_i  \sigma^n_{j}
\left(
{^{A~1~~1}_{2~M~N}}
\right)
\epsilon_{A}~,
\label{8}
\eeq
where
\bea
\epsilon_{A} &=& \sum_{L=J}  {\sqrt{32 (\overline{n}^2 - 4)\over 15
 (\overline{n}^2 - 1)}}
b^{MN}_{JL}(\eta)  Y^{2~LJ}_{m~MN}(g) +
\sum_{|J-L|=1}  {\sqrt{32 (\overline{n}^2 - 4)\over 5}}
c^{MN}_{JL}(\eta)  Y^{2~LJ}_{m~MN}(g) + \nonumber \\
&& + \sum_{|J-L|=2}  {\sqrt{32 \over 5}}
d^{MN}_{JL}(\eta)  Y^{2~LJ}_{m~MN}(g),
\label{9}
\eea
with $\sigma^m_i$ described below, the coefficients $a^{MN}_{J},\ldots ,
d^{MN}_{JL}$ depend only on the conformal
time, $\left(
{^{L~J~~~j}_{N~M~m}}\right)$ represent 3-J symbols or Clebsh-Gordon
coefficients and $\overline{n} = J + L + 1$.

The massive vector field
\begin{equation}
\mbox{\boldmath $A$} = A_m^{ab}\omega_{s}^m {\cal T}_{ab} =
 A_m^{ab} \sigma^m_i \omega^i {\cal T}_{ab},
\label{vec}
\end{equation}
where $\omega_{s}^m$ denote the one-forms in a spherical basis with
$m=0,\pm 1$, ${\cal T}_{ab}$ are the $SO(3)$ group generators and
\begin{equation}
\sigma^m_i =
\left[
 \begin{array}{ccc}
-\frac{i}{\sqrt{2}} & \frac{1}{\sqrt{2}} & 0 \\
\frac{i}{\sqrt{2}} & \frac{1}{\sqrt{2}} & 0 \\
0 & 0 & i
\end{array}
\right],
\label{eq:1/13/b}
\end{equation}
can be expanded in spin-1 hyperspherical harmonics as  \cite{tt}:
\bea
A_0(\eta,x^j) &=& \sum_{JMN} \alpha^{abJM}_{\phantom{^{JM}} N}(\eta)
{\cal
D}^{J{\phantom J}M}_{{\phantom J}N} (g) {\cal T}_{ab}
= 0 + \sum_{J'M'N'} \alpha^{abJ'M'}_{\phantom{^{J'M'}} N'}(\eta) {\cal
D}^{J'{\phantom J'}M'}_{{\phantom J'}N'} (g) {\cal T}_{ab},
\label{11a} \\
A_i(\eta,x^j) &=& \sum_{LJNM} \beta^{abMN}_{LJ} (\eta)
{}~Y^{1LJ}_{mNM}(g) \sigma^m_i~{\cal T}_{ab} \nonumber \\
 & = & {1 \over 2}\left[1 + \sqrt{{{2\overline{\alpha}}
\over
{3\pi}}} \chi(\eta)\right]\epsilon_{aib}{\cal T}_{ab}
+ \sum_{L'J'N'M'} \beta^{abM'N'}_{L'J'} (\eta)
{}~Y^{1L'J'}_{mN'M'}(g) \sigma^m_i~{\cal T}_{ab},
\label{11b}
\eea
where $\overline{\alpha} = e^2/4\pi$ and
$A_0$ is a scalar on each fixed time
hypersurface, such that it can be expanded in scalar harmonics
${\cal D}^{J{\phantom J}M}_{{\phantom J}N}(g)$.
The coordinates $x^i$ are written as an element of
$SU(2)$. The expansion of $A_i$ is performed in terms of the spin-1
spinor
hyperspherical harmonics, $Y^{1~LJ}_{m~MN}(g)$. Longitudinal and
transversal harmonics correspond to $L = J$ and $L-J = \pm 1$,
respectively.
The $\alpha^{abJM}_{\phantom{^{JM}} N}(\eta)$ and $\beta^{abMN}_{LJ}
(\eta)$
are time-dependent functions and each one identifies a spin-1 mode
from the $A_0$ and $A_i$ components, respectively.
The expressions in the first equality, eqs.
(\ref{11a}) and (\ref{11b}), represent the general expansion.
In the $U(1)$ model, we choose the homogeneous modes of the spin-1 field
spatial
components to be identically zero. In this way, the homogenous modes
will
produce field strenght configurations compatible with FRW geometries,
i.e.
a diagonal energy-momentum tensor, vanishing for the U(1) case.
In addition,  the r.h.s. of
the same expressions
correspond to a decomposition of the expansion in homogeneous (first
term) and
inhomogeneous modes for the non-Abelian case.
There, we use a $SO(4)$-symmetric Ansatz for the homogeneous modes
of the vector field which is compatible with the FRW geometry \cite{pp}.
For the case of a $SO(3)$ global symmetry, we have, for the homogeneous
modes
\begin{equation}
A_0^{ab}(\eta) = 0,
\end{equation}
\begin{equation}
\mbox{\boldmath $A$}_i(\eta) =  {1 \over 2}\left[1 +
 \sqrt{{{2\overline{\alpha}}\over
{3\pi}}} \chi(\eta)\right]\epsilon_{bic}{\cal T}_{bc},
\end{equation}
where $\chi(\eta)$ is time-dependent scalar function.
The idea underlying this Ansatz for the non-Abelian
spin-1 field consists in defining an homorphism from the
isotropy group $SO(3)$ to the gauge group.
This homomorphism defines the internal  transformation
which, for the symmetric fields,
compensates the action of a given $SO(3)$ space rotation.
Hence, the above form
for the gauge field, where the $A_0$ component is taken
to be identically zero. By imposing the
above mentioned symmetry conditions, we obtain a
one-dimensional mechanical-type model depending only on time \cite{pp}.
The resulting
one-dimensional model has some residual
symmetries from the ones of the full
four-dimensional theory. In particular, the invariance
under general coordinate transformations in four
dimensions leads
to an invariance under arbitrary
time--reparametrizations. However, due to our choice
of $SO(4)$--symmetry conditions for the spin-1 field,
none of the local internal
symmetries survive as all the available
internal transformations are required to
cancel out the action of a given $SO(3)$ space rotation.

Let us now turn to our model with a massive vector field. From actions
(\ref{1})
and  (\ref{4})
one can work out the effective Hamiltonian density. Upon substitution of
the expansions (\ref{2})-(\ref{11b}) and after integrating over $S^3$,
the canonical conjugate momenta of the dynamical variables are found to
be
\beq
\mbox{\boldmath $\pi$}_{a} = {\partial {\cal L}^{{\rm
eff}}\over\partial\dot a}
 = -\frac{\dot a}{N}~,~~
\mbox{\boldmath $\pi$}_{\chi} =
{\partial {\cal L}^{{\rm eff}}\over \partial \dot\chi} =
\frac{\dot\chi}{N}~.
\label{13a}
\eeq

\beq
\mbox{\boldmath $\pi$}_{\beta^{abNM}_{LJ}} = {\partial {\cal L}^{{\rm
eff}}
 \over\partial\dot\beta^{abNM}_{LJ}}
= \frac{\dot\beta^{abNM}_{LJ}}{2N\pi\overline{\alpha}}~,
\label{13c}
\eeq
\beq
\mbox{\boldmath $\pi$}_{\beta^{abNM}_{JJ}} = {\partial {\cal L}^{{\rm
eff}}
 \over\partial\dot\beta^{abNM}_{JJ}}
= \frac{\dot\beta^{abNM}_{JJ}}{2N\pi\overline{\alpha}} -
\frac{1}{2\pi\overline{\alpha}}\frac{a}{N}
\alpha^{abJM}_{\phantom{^{JM}} N}
(-1)^{2J} \sqrt{{16\pi^2 J(J+1)\over 2J+1}},
\label{13b}
\eeq
where ${\cal L}^{{\rm eff}}$ denotes the effective Lagrangian density
associated with (\ref{1}), which we omit here,  and the dots represent
derivatives with respect to
the conformal time.
To second order in the coefficients of the
expansions and all orders in $a$, one finds that most of the
gravitational
degrees of freedom are gauge type and, as such, the wave function
cannot depend on them;
furthermore, we shall consider the gravitons in the ground state.
For the case of $SO(3)$ global symmetry, dropping the primes,
it  follows that the effective Hamiltonian density reads
(where for the Abelian case\footnote{For other presentations of an
Hamiltonian formulation of systems involving the Proca field, see e.g.
Refs. \cite{pilati,roman}.}, one disregards the last four terms):
\bea
{\cal H}^{{\rm eff}} &=& -\frac{1}{ 2M_P^2} \mbox{\boldmath $\pi$}^2_a +
 M_P^2\left( - a^2 +
{4\Lambda\over 9\pi }~a^4
\right) \nonumber
\\
&& + \sum_{J,L}~{4\over 3\pi} m^2 a^2 \beta^{abNM}_{LJ}
{}~\beta^{abLJ}_{NM} \nonumber \\
&& + \sum_{\vert J-L\vert =1} \left[ \overline{\alpha}\pi~
\mbox{\boldmath
 $\pi$}_{\beta^{abLJ}_{NM}}
\mbox{\boldmath $\pi$}_{\beta^{abLJ}_{NM}} + \beta^{abNM}_{LJ}
\beta^{abLJ}_{NM}
 (L+J+1)^2\right]
\nonumber \\
&& + \sum_J \left\{
\overline{\alpha}\pi + \left[(-1)^{4J} \left({16\pi^2 J(J+1)\over
2J+1}\right)
 ~{3\pi
\over 4m^2}\right]~~
\left[ 1 + {1\over a(t)}\right] \right\}\mbox{\boldmath
 $\pi$}_{\beta^{abJJ}_{NM}}
\mbox{\boldmath $\pi$}_{\beta^{abJJ}_{NM}} \nonumber \\
&& + \mbox{\boldmath $\pi$}^2_{\chi} + {\overline{\alpha}\over 3\pi}
\left[
 \chi^2  - {3\pi\over
2\overline{\alpha}}\right]^2  \nonumber \\
&& + \sum_{J,L} {4\over \overline{\alpha}\pi}~\left[ 1 +
 \sqrt{{2\overline{\alpha}\over 3\pi}} \chi
\right]^2~~\beta^{abNM}_{LJ} ~\beta^{abLJ}_{NM} \nonumber \\
&& + 4\pi a^2 m^2~\left[ 1 + \sqrt{{2\overline{\alpha}\over
3\pi}}\chi\right]^2,
\label{12}
\eea
where, from now on, for brevity, we shall colectively  denote (J,L,M,N)
by J,
being implicit the difference between longitudinal and transversal
modes as well as the sum over contracted $SO(3)$ group indexes.
In the Appendix, neglecting the expansions (\ref{7})-(\ref{9}), we
present the complete effective Hamiltonian
for the perturbed non-Abelian model.

The Hamiltonian constraint, ${\cal H}^{{\rm eff}} = 0$, gives origin to
the
Wheeler-DeWitt equation after promoting the canonical conjugate momenta
(\ref{13a}),(\ref{13b}) into operators:
\beq
\mbox{\boldmath $\pi$}_{a} = -i{\partial\over\partial
a}~,\mbox{\boldmath
 $\pi$}_{\chi} =
-i{\partial\over\partial
\chi}~,~~ \mbox{\boldmath $\pi$}_{\beta^{abNM}_{LJ(JJ)}} =
 -i{\partial\over\partial
\beta^{abLJ(JJ)}_{NM}}~,
{}~~\mbox{\boldmath $\pi$}_{a}^2 = - a^{-P}{\partial\over\partial a}~
\left( a^P{\partial\over\partial a}\right)~,
\label{14}
\eeq
where in (\ref{14}) the last substitution parametrizes the operator
order
ambiguity with $P$ being a real constant.
The Wheeler-DeWitt equation is obtained imposing that the Hamiltonian
operator
annihilates the wave function $\Psi
 [a,\chi,\beta^{abNM}_{LJ},\beta^{abNM}_{JJ}]$.

Before we proceed in discussing the way the decoherence process takes
place
within the framework of our model let us comment on some features of the
Hamiltonian density (\ref{12}). The first line in (\ref{12}) correponds
to the
contribution from gravitational degrees of freedom in their ground
state. The
third line is associated with the transversal modes $(J-L = \pm 1)$ of
the
spin-1 field. The fourth line corresponds, on its hand, to the
contribution of
the longitudinal modes $(J=L)$. Notice that the presence of the mass
term in
the model entangles all modes of the spin-1 fields (longitudinal and
transversal) with the metric as exhibited in the second line in
(\ref{12}). The
mass of the vector field is also present in the longitudinal
kinetic part,
where there is also a  term proportional to $a(t)^{-1}$. In the fifth
line, in (\ref{12}), one has the kinetic piece of the homogeneous part
of the
spin-1 field as well as the quartic potential, typical from the
dimensional
compactification procedure for treating the cosmological problem of
coupling
gravity with Yang-Mills and vector fields
\cite{lll},\cite{nn}--\cite{ss1}. The
coupling between the homogeneous, and all the inhomogeneous modes of the
vector
field are shown in the sixth line. Finally in the seventh line in
(\ref{12})
one has the coupling of the gravitational ground state mode to the
homogeneous
part of the mass term of the spin-1 field.
Thus, our system can be regarded as a set of coupled harmonic
oscillators
(gravitational and spin-1 field) where the kinetic term of the
longitudinal
modes has a  coefficient proportional to $a(t)^{-1}$.

In what follows, we shall compare our models with the
ones discussed in the literature when studying
correlations between coordinates and momenta and
the decoherence process.

\section{Decoherence and Back-Reaction Processes in the Presence of
 Massive Abelian Vector Fields}

\indent

In this Section, we discuss the process of decoherence
in a closed FRW model with a massive vector field
and global $U(1)$ symmetry and its relation to the
conditions which support correlations between coordinates and canonical
momenta. The validity of the semi-classical Einstein equations, i.e. the
the so-called back-reaction problem, is also discussed. As explained in
the previous sections,
we take the homogeneous modes of the spatial
components of the vector field to be zero,
whereas the environment is identified with the inhomogeneous modes,
i.e. the $\beta_{J}$-functions (see
 eq. (\ref{12a}) below).
Therefore, our minisuperspace  will be one-dimensional and the
FRW scale factor will
correspond to the physical observable about which predictions can be
made.

We rewrite the Wheeler-DeWitt equation (\ref{12}),(\ref{14}) for the
$U(1)$ case as
\bea
{\cal H}^{{\rm eff}} \Psi[a,\mbox{\boldmath $A$}_{\mu}]
&=& \left[\frac{1}{ 2M_P^2}
\frac{\partial^2}{\partial a^2}
-
M_P^2\left(  a^2 - {4\Lambda\over 9\pi }~a^4
\right)\right. \nonumber\\
&& \left. - \sum_J
f_J(a)\left[
\frac{\partial^2}{\partial (\beta^{J})^2}
 -
\Omega^2_J(a)~ \beta_{J}~\beta_{J} \right] \right]\Psi[a,\mbox{\boldmath
 $A$}_{\mu}] = 0 ,
\label{12a}
\eea
where
\begin{equation}
f_J(a) = \left\{ \begin{array}{cll}
\overline{\alpha}\pi & {\rm if} & J-L=\pm 1 \\
\overline{\alpha}\pi + \left[(-1)^{4J} 48\pi^3 J(J+1)/( (2J+1) 4 m^2)
\right]~~
\left[ 1 + (1/a(t))\right] & {\rm if} & J=L
\end{array}  \right. ,
\label{12b}
\end{equation}
and
\begin{equation}
\Omega^2_J(a) = \left\{ \begin{array}{cll}
[(4m^2 a^2/3\pi) + (L+J+1)^2]/
\overline{\alpha}\pi & {\rm if} & J-L=\pm 1 \\
 \frac{(4m^2 a^2/3\pi) + (2J + 1)^2 +1}{ \overline{\alpha}\pi
+ \left[(-1)^{4J} 48\pi^3 J(J+1)/( (2J+1) 4 m^2) \right]~~
\left[ 1 + (1/a(t))\right]}
 & {\rm if} & J=L
\end{array}  \right. ,
\label{12c}
\end{equation}
following the notation of Ref. \cite{ggg} and setting the ordering
ambiguity
 factor to
$P=0$. Our massive $U(1)$ model and  the
$k=+1$ FRW minisuperspace model with a massive conformally coupled
scalar field discussed in Ref. \cite{ggg}
share some similar features. Indeed, up to different constant
coefficients and
the $1 + 1/a$ factor, the Hamiltonian (\ref{12}) is equivalent to the
one of a
massive conformally coupled scalar field in a closed FRW background.
As far as the $1 + 1/a$ factor is concerned,
if one considers expanding solutions, then the condition
 $1 + 1/a \rightarrow 1$ will be rapidly satisfied and,
hence, we can apply the framework used in \cite{ggg} and draw similar
conclusions.
However, the case of contracting solutions as well as the interference
between
the two WKB branches have to be addressed  differently.

A solution of (\ref{12a})--(\ref{12c}) which corresponds to a
``classical'' behaviour
of the $a$-variable  on some region of minisuperspace will have an
oscillatory WKB form (\ref{wkbinf})  as
\begin{equation}
\Psi_{(n)} [a,\mbox{\boldmath $A$}_\mu] = e^{iM_P^2 S_{(n)}(a)}
C_{(n)}(a)
 \psi_{(n)}
(a,\mbox{\boldmath $A$}_\mu)~.
\label{WKB}
\end{equation}
After expanding the functions in (\ref{WKB}) in
powers of $M_P$ and using (\ref{12a})--(\ref{12c}) one
finds that the lowest order action, $S_0$, satisfies the Hamilton-Jacobi
equation,
\beq
-\frac{1}{2}S_0^{\prime 2} + V(a) = 0~,
\label{HJ}
\eeq
 where
$V(a) =   - a^2 +  {4\Lambda\over 9\pi }~a^4$ and the prime denotes
derivative with respect to $a$. From (\ref{12a})--(\ref{12c}) we see
that the different modes do not interact among themselves. Thus, the
wave
function  $\psi_{(n)} (a,\mbox{\boldmath $A$}_\mu)$ can be
factorized as
\begin{equation}
\psi_{(n)} (a,\mbox{\boldmath $A$}_\mu) \equiv \psi_{(n)} (a,\left\{
\beta_J
 \right\} )
= \prod_J \psi_{(n)J} (\eta, \beta_{J})~.
\label{use}
\end{equation}
Defining the  WKB time as
\beq
\frac{d}{d\eta} = \frac{\partial S}{\partial a}\frac{d}{d a}~,
\eeq
one obtains the Schr\"odinger equation satisfied by each
wave function $\psi_{(n)J}(\eta, \beta_{J})$:
\beq
\frac{1}{2} f_J(a) \left(-\frac{\partial^2}{\partial (\beta^{J})^2}
+ \Omega^2_J (\beta^{J})^2 \right)\psi_{(n)J}(\eta, \beta_{J}) = i
 \frac{d}{d~\eta}
\psi_{(n)J}(\eta, \beta_{J})~.
\label{Schdg}
\eeq
We stress that
$\psi_{(n)J}(\eta, \beta_{J})$ is actually
$\psi_{(n)J}[a(\eta), \beta_{J}]$, dependent on the ``classical''
physical observable and $\beta_{J}$ as well.
 From (\ref{Schdg}), we can calculate, say,
$\psi_{(n)J}(\eta, \beta_{J})$ in $\eta^\prime$ given the value of
$\psi_{(n)J}(\eta, \beta_{J})$ at $\eta^{\prime\prime} < \eta^{\prime}$.

In order to make predictions concerning the behaviour of $a$,
one uses a coarse-grained description of the system
working out the reduced density matrix associated with the
WKB wave function of the form
(\ref{WKB}) to obtain \cite{cm,cm1}
\begin{equation}
\rho_R = \sum_{n,n'} e^{iM_P^2[S_{(n)}(a_1) - S_{(n')}(a_2)]}
C_{(n)}(a_1)C_{(n')}(a_2) {\cal I}_{n,n'}(a_2,a_1)~,
\label{red}
\end{equation}
where
\begin{equation}
{\cal I}_{n,n'}(a_2,a_1) \equiv  \int \psi^*_{(n')} (a_2,\mbox{\boldmath
 $A$}_{\mu}) \psi_{(n)}
(a_1,\mbox{\boldmath $A$}_{\mu}) d[\mbox{\boldmath $A$}_{\mu}]
= \Pi_J \int \psi^*_{J(n')} (a_2,\beta_J) \psi_{J(n)}
(a_1,\beta_J) [ d \beta_J]~.
\label{infl}
\end{equation}
The subindex $(n)$ labels the WKB branches. The term
${\cal I}_{n,n'}(a_2,a_1)$ describes the influence of the environment
on the system.
Notice that in (\ref{infl}) {\em all} modes must be included.

The analysis of correlations between minisuperspace coordinate and
momenta is,
in quantum cosmology,
usually discussed using the Wigner function criterion \cite{bb,ts1,ts2}:
A strong sharp peak is likely to be located close to  a classical
trajectory defined by  the
Hamiltonian-Jacobi equation plus quantum-corrections. However,
the Wigner function associated with the reduced density matrix
(\ref{red}) does not have a single sharp peak even for a pure WKB
function as
(\ref{WKB}) or a linear combination of them (cf. Refs.
\cite{llo,new1,new2,new3}). Nevertheless,
this problem can be
overcome through the environment interaction with  the ``observed''
system \cite{llo}.
As explained in the Introduction, such interaction is at the origin of
the loss of quantum-coherence or
decoherence between different classical trajectories, i.e. WKB branches.
More precisely, correlations between coordinates and
momenta must be analysed within each classical branch ($n=n'$).
 This can be done by looking at the reduced density matrix
associated with it or the corresponding Wigner functional:
\begin{equation}
F_{W(n)} (a,\pi_a) = \int_{-\infty}^{+\infty} d\Delta
[S'_{(n)}(a_1) S'_{(n)}(a_2)]^{-\frac{1}{2}} e^{-2i\pi_a\Delta}
e^{iM^2_P[S_{(n)}(a_1) - S_{(n)}(a_2)]} {\cal I}_{n,n}(a_2,a_1)
\label{wigf}
\end{equation}
where $\Delta = (a_1 - a_2)/2$.
A correlation among variables will correspond to a strong peak
about a classical trajectory in the phase space.
Thus, there exists an important relation between correlation
and decoherence as one needs the
latter, i.e. fairly small off-diagonal terms in
(\ref{red}) such that  quantum interference
between alternative histories is negligible
($ {\cal I}_{n,n'} \propto \delta_{n,n'}$), in order to obtain the
former.
Hence, the decoherence process is rather crucial as it is only when
the decoherence between different
WKB branches is sucessful that correlations may be properly predicted.

Besides the decoherence between different WKB branches,
the environment interaction also affects the correlations within
a classically decohered branch; this is explicit in the
functional
${\cal I}_{n,n}(a_2,a_1)$ in eq. (\ref{wigf}). As pointed out by
Zurek \cite{aaa},
the environment degrees of freedom {\em continuously measure}
the physical observables and
this interaction not only suppresses the off-diagonal ($n\neq n'$) terms
in
(\ref{red}) and (\ref{infl}), but also
induces a ``localizing'' effect on
the classical variables within each WKB branch. This corresponds to
the back-reaction from the environment
on the semi-classical evolution of the system. In particular,
${\cal I}_{n,n}(a_2,a_1)$ will be damped for $|a_2 - a_1|\gg 1$ and the
reduced density matrix associated with (\ref{wigf}) will
be {\em diagonal} with respect to
$a$. The sharpness and position of the peak
will be determined by the behaviour of
${\cal I}_{n,n}(a_2,a_1)$. Furthermore, it has been shown in Refs.
\cite{bb,cm} that the localization
effect inside a classical branch is much more efficient than the
decoherence
between diferent WKB branches. It has been also remarked in Refs.
\cite{ggg,ggga} that, if the conditions
for achieving an effective localization (and diagonalization)
of (\ref{wigf}) are met, then the
interference between the different WKB branches  is {\em also}
highly supressed. Actually, the functional ${\cal I}_{n,n'}(a,a)$ has
been usually identified as a measure of the
decoherence between two different WKB histories, characterized by
the parameters $(n)$ and $(n')$;
an heuristic argument in support of that view was presented
in Ref. \cite{ggga}.

Before proceeding, we point out that using a
Gaussian Ansatz  for the environment state as
\begin{equation}
\psi_J (a,\beta_J) = D_J(t) e^{i\gamma_J(t) - B_J(t)\beta_J^2}~,
\label{gauss}
\end{equation}
for each $\psi_J (\eta, \beta_J)$, where
$D_J,\gamma_J$ are real, $B_J(t) = B_{rJ} + iB_{iJ}$, where $B_{rJ}$ and
$B_{iJ}$ are also real and  $B_{rJ} > 0$, and  imposing the
normalization
 condition
\beq
 \int \psi^*_J (t,[\beta_J]) \psi_J
(t,[\beta_J])  d \beta_J = 1~,
\label{norm}
\eeq
 the general form of
${\cal I}_{nn'}(a,a')$ for any mode is given by
\beq
{\cal I}_{(nn')J}(a,a') = \exp \left[ i \left( D_{(n')J}(a') -
D_{(n)J}(a)
 \right) \right]
\left[ \frac{4 B_{(n')rJ}(a') B_{(n)rJ}(a)}{ \left(  B_{(n')J}(a') +
 B_{(n)J}(a) \right)^2}
 \right]^{1/4} .
\label{gusinn}
\eeq
The analysis of the  decoherence between different WKB histories
and correlations via
the Wigner function (with ${\cal I}_{(n,n)J}(a,a')$) requires
the functions $B$ and $D$ to be  found explicitly.
The requirements for sucessfull diagonalization  and ``localization''
were {\em generally} established  and  discussed in Refs.
\cite{ggg,ggga}. From
 the assumption that
$|a_1 - a_2| \ll 1$  and the Gaussian Ansatz (\ref{gauss}), (\ref{norm})
for $\psi_J (a,\beta_J)$, the above mentioned conditions read:
\begin{equation}
\left(\sum_J \frac{B^{\prime}_{iJ}}{2B_{iJ}}\right)^2 \ll \sum_J
\frac{|B^{\prime}_{J}|^2}{4B_{rJ}^2}~,
\label{adyab}
\end{equation}
\begin{equation}
2M_P^4 \left[V(a)  + \frac{1}{M_P^2} \sum_J \left(  \frac{B_{rJ}^2 +
 B_{iJ}^2}{2B_rJ} +
\frac{\Omega^2_J}{8 B_{rJ}}
 \right)
\right]  \gg
\sum_J
\frac{|B^{\prime}_{J}|^2}{4B_{rJ}^2}~,
\label{Scorr}
\end{equation}
\begin{equation}
\sum_J
\frac{|B^{\prime}_{J}|^2 \overline{a}^2}{4B_{rJ}^2} \gg 1~,
\label{Sdecoh}
\end{equation}
where  $\overline{a} = (a_1 + a_2)/2$ (the sums in these expressions
and previously
related ones have an implicit factor arising from the degeneracy  of the
$J,L,M,N$ mode). Expressions (\ref{adyab})--(\ref{Sdecoh})
are usually referred to as
{\em adiabaticity}, {\em strong decoherence} and
{\em strong correlation} conditions, repectively. Notice that the
results
(\ref{adyab})--(\ref{Sdecoh}) arise directly from
(\ref{infl}),(\ref{wigf})
and the Gaussian Ansatz for the state of the environment (\ref{gauss}).
Henceforth, the validity of these conditions has to be
analysed using the quantities and parameters relevant to our particular
models.
The possibility that the
massive spin-1 field models give rise to new conditions for the process
of
correlation and  decoherence has to be properly considered.
The adiabaticity condition warrants the validity of the zero-th order
WKB
evolution as its violation implies that the semiclassical Einstein
equations
are not valid due to contributions of high-order in the phase of
${\cal I}_{n,n}(a_2,a_1)$. On the other hand, (\ref{Scorr}) reflects the
fact
that the peak in the Wigner function (shifted away from the expected
classical trajectory by interaction with the environment) is sharp
as far as the center of the peak is large
when compared to the spread. Finally, expression (\ref{Sdecoh})
translates
the condition of strong decoherence corresponding effectivelly to the
requeriment of diagonalization of the
reduced density matrix associated with (\ref{wigf}).
It is important to mention that usually a compromise
between decoherence and correlation is needed since
if the later is too strong, then the peak in the Wigner function is
actually
broaden \cite{hl}.

Let us now address the issues of decoherence, correlations
and back-reaction in our model with massive Abelian vector fields.
Firstly, we shall assume that  the decoherence between the two
different WKB histories has
occured sucessfully and
consider the correlation and localization effects within a classical
branch.
Afterwards, we shall comment on
the decoherence between
different WKB branches.
The inclusion of transversal as well as
longitudinal modes in eqs. (\ref{red}),(\ref{infl}),
(\ref{adyab})--(\ref{Sdecoh})  give rise to some
difficulties for our $U(1)$ and $SO(3)$ models as far as the
longitudinal modes are concerned.
This will be discussed in the following. In particular, notice that
eq. (30)--(32) and then (36)--(38) involve considering {\em all}
modes, longitudinal and transversal. The same will
apply to other equations as will be pointed out explicitely.

We start by considering the correlation and localization effects within
each
classical branch for the case of tranversal modes\footnote{The
computations corresponding to each of the two WKB branches
(expanding and contracting) are the same in the two cases
($n=\pm 1$, say).}  ($J\neq L$). One can easily verify that up
to a re-scaling of the
$\beta_{JL}$-modes by a factor of $\overline{\alpha}\pi$ (and conversly
for the
corresponding canonical momenta), this part of the model is equivalent
to
the one with a massive
conformally coupled scalar field (cf. Refs. \cite{ggg,cm1}).
Substituting (\ref{gauss}) into the Schr\"odinger equation
(\ref{Schdg}), we get
the following equations
\begin{eqnarray}
\dot{\gamma}_J & = & - f_J(a)  B_{rJ}~, \label{eq1} \\
\dot{B}_J & = & i f_J(a) [-2(B_J^2 - \Omega^2_J/4)]~, \label{eq2}
\end{eqnarray}
for $D_J=\pi^{-1/4}(2B_{rJ})^{1/4}$. With the above  mentioned
re-scaling,
eq. (\ref{eq2}) can be linearized via the choice
$B_J=-i\dot{\varphi}_J/2 \varphi_J$, to yield
\beq
\ddot{\varphi}_J + \Omega^2_J \varphi_J = 0~.
\label{fakekg}
\eeq

The initial state of the environment is associated
with a particular choice of initial conditions when  solving the
preceding
 equation. In our
present case, the Hamiltonian eq. (\ref{12a}) corresponds to a set of
harmonic oscillators with a variable time--dependent mass.
A convenient vacuum state can be defined  assuming there
 exists an adiabatic zone such that the
classical evolution emerging from the Hamilton-Jacobi equation satisfies
the condition $\dot{a}/a \rightarrow 0$ for large $a$.
This requires that $V(a)$ be quadratic in $a$, meaning that models with
non-vanishing cosmological constant do not satisfy this
adiabaticity condition as, in this case, $V(a) \sim O(a^4)$ for large
$a$.
Notice that the Hamilton-Jacobi equation
for a generic $V(a)$ has
real solutions for $a>a_0$ only if $a_0$ is a single zero of $V(a)$
for $V(a) > 0$.
Assuming a vanishing cosmological constant, one can identify a vacuum
for  the adiabatic {\em out} regime ($a\gg 1$), being
the out-modes of the form \cite{ggg,ggga,cm,cm1}
\beq
\varphi_j^{out} = (2\Omega_J)^{-1/2} \exp\left[-i\int^{\eta}
 \Omega_J(\eta^{\prime})
d\eta^\prime\right],
\label{outmodes}
\eeq
which diagonalize asymptotically the Hamiltonian for large values of
$a$.
For small values of $a$, a preferred initial state (an {\em in} vacuum
state)
may in some cases  also be defined as the one
which diagonalizes the Hamiltonian for $a>a_0$ (in our model for
$\Lambda=0 $ and for other commonly used, $d a/ d \eta \ll 1$, for small
values of $a$).
The relation between the {\em in} and {\em out} modes is given by
the Bogolubov transformation
\beq
\varphi_J^{in} = \hat{\alpha}_J \varphi_J^{out} + \hat{\beta}_J
 (\varphi_J^{out})^*.
\label{bogo}
\eeq
The  $\hat{\alpha}_J,\hat{\beta}_J$ are designated as Bogolubov
coefficients
and any particular choice for these determine different vacuum states
for
 $\varphi$.
One obtains for a general potential $V(a)$
inducing an adiabatic behaviour  \cite{cm}
\beq
\hat{\alpha}_J \simeq 1~;~\hat{\beta}_J \simeq \frac{i}{2} \exp
\left\{ -\frac{1}{2} \pi [m^2 a_0 V^\prime (a_0)]^{-1/2} (d_J^2 + m^2
 a_0^2)\right\},
\label{bogocof}
\eeq
where the coefficient $d_J$ denotes the degeneracy associated to
$J,L,M,N$.
Notice that for the case of a quadratic  $V(a)$ one obtains
$\hat{\beta}_J = 0$.
If the condition  $\hat{\alpha}_J \simeq 1 \gg \hat{\beta}_J \simeq 0$
is
satisfied, which occurs when the evolution is
indeed adiabatic, then such quantum state is usually identified
as adiabatic vaccum and holds during all the evolution.
Notice again that, in the case of a non-vanishing cosmological constant,
the adiabaticity condition ($\dot{\Omega}_J/\Omega_J^2
\ll 1$) cannot be satisfied.

Since the massive spin-1 transversal modes behave effectivelly
as conformally coupled massive scalar fields, it comes as no surprise
that the
{\em adiabaticity}, {\em strong decoherence} and
{\em strong correlation} conditions
(\ref{adyab})--(\ref{Sdecoh}) are indeed satisfied {\em restricted}
to those environment modes.
Notice then that
\beq
\sum_J
\frac{|B^{\prime}_{J}|^2}{4B_{rJ}^2}  = \sum_J \left[
\frac{1}{4} \left( \frac{\Omega_J^\prime}{\Omega_J }\right)^2 +
|\hat{\beta}_J|^2\left(4\Omega_J^2\frac{1}{\dot{a}}\right)^2 \right].
\label{decohout}
\eeq
Within the adiabaticy evolution requirement, we take a quadratic $V(a)$
and
 hence
the first term in (\ref{decohout}) will be dominant. Using eqs. (23) and
(24),
we see that the sum in
eq. (\ref{decohout}) implies (cf. Ref. \cite{ggg}) that the
 strong decoherence condition (\ref{Sdecoh}) corresponds
asymptotically to
\beq
m^3 \overline{a}^3 \gg 1,
\label{dechoout1}
\eeq
while the strong correlation condition (\ref{Scorr}) reduces
asymptotically
to
\beq
|V(\overline{a})| \gg m^{-1}\overline{a},
\label{scorrout}
\eeq
for large $a$. As long as we restrict ourselves to assess the
decoherence process and correlation analysis for the case
of an {\em expanding} solution (as it is for  the cases
studied in most of the literature) the conditions (46),(47) are valid.
In fact, that seems to be the right interpretation when
explaning the
quantum to classical transition of our Universe. As one can see,
the mass of the of the Abelian vector field provides a scale at
 which the process
of decoherence and the analysis of correlations take place. Indeed, for
fairly small or negligible mass the process of
decoherence is not completely achieved, which is consistent with the
decoupling
between the gravitational and the Yang-Mills field ($m=0$) [12].
The adiabaticity condition implies that the growth
of $V(a)$ must be  slowlier than that of a quartic potential.
For a quadratic $V(a)$ this is immediate.

One could, for instance, consider
an {\em ad hoc} potential from the start \cite{ggg}, although it would
remain
to be verified if such a potential would satisfy conditions
(\ref{adyab})--(\ref{Sdecoh}) and could be derived from a realistic
action
(with anisotropy or even higher curvature terms).
On its turn, the presence of a cosmological
constant induces divergences
in the decoherence factor (\ref{decohout}) in addition to the ones from
the back-reaction factor.
If the latter were expected to correspond to the zero- point energy of
the fields, the former
may only be cured via the introduction of a fundamental cut-off since
it cannot be removed by
standard renormalization procedure \cite{ggg,ggga}. However, that seems
rather unsatisfactory from the physical view point. Considering
the environment composed by modes whose physical wavelength
is larger than the horizon is in disagreement with the expectation that
the
environment consists of small wavelenght fluctuations.

Let us now return to the decoherence between different WKB branches. The
quantity to analyse is ${\cal I}_{nn'}(a,a)$ (see above remarks and Ref.
 \cite{ggga}).
We suppose once again that our minisuperspace model has an adiabatic
{\em out} region for large values of $a$. Then, up to second order
for the adiabatic limit and $\hat{\beta}$-Bogolubov
coefficients, one can find, in the case of the transversal modes of the
massive Abelian field \cite{ggga,cm,cm1}:
\begin{eqnarray}
{\cal I}_{nn'}(a,a)
& \simeq & \exp \left\{ \sum_J\left[
 -  \frac{1}{4}\left( |\hat{\beta}_{nJ}|^2 + |\hat{\beta}_{n'J}|^2 +
2~\hat{\beta}_{nJ}~\hat{\beta}_{n'J}
\cos\left[2\int^\eta (\Omega_{nJ} - \Omega_{n'J})  \right] \right)
\right. \right.
\nonumber \\
& + & \left. \left.
\frac{1}{16} \frac{(\dot{\Omega}_{nJ} -
\dot{\Omega}_{n'J})^2}{\Omega^4_{J}}
\right]\right\},
\label{Inn}
\end{eqnarray}
where a sum over the environment transversal modes is understood
(the longitudinal modes will be treated  in the next paragraph).
The case of a quadratic $V(a)$
imposes $\hat{\beta}_{JL} = 0$ and, as mentioned previously, this means
that the quantum state of the environment evolves as an adiabatic
vacuum.
In that case, for $\Omega_n = - \Omega_{n'}$ we have
\beq
{\cal I}_{nn'}(a,a) \simeq \exp \left\{ - \frac{1}{16} \sum_J
 \frac{(\dot{\Omega}_{nJ})^2}{\Omega^4_{J}}
\right\}~,
\label{Inn1}
\eeq
and, since $\frac{\dot{\Omega}_{nJ}}{\Omega^2_{J}} \ll 1$, these terms
are effectively destroyed
for the two WKB histories. Another possibility is to consider that the
Universe has undergone
a static or quasi-static period for large $a$ such that
$\dot{\Omega} \sim 0$. In that case, the framework established above
for the $\varphi_J$ modes can be used for the
environment and a natural {\em in} vacuum state can be defined,
which is not equivalent to the {\em out}  vacuum \cite{cm}.
Associated particle creation takes place and, as a
consequence, decoherence occurs and the relation between
the amount of interference that is suppressed
and the number of particles created is given by
\beq
{\cal I}_{nn'}(a,a) \simeq \exp \left\{ -4 \sum_J d_J^2
\hat{\beta}^2_J\cos^2\left[2\int^\eta \Omega_{nJ}\right]
\right\}.
\label{Inn2}
\eeq

Finally, one has to consider the longitudinal modes ($J=L$).
Surely, they have to be included in eqs. (45) and (48)--(50) if the
results
(46),(47) and loss of quantum-interference among the expanding
and contracting
WKB branches are to be extended to a mode-complete massive Abelian
vector field.
As far as expanding solutions are concerned, namely the ones for
which $a$ becomes much larger than $1$ sufficiently fast, the previous
remarks for the transversal modes can be extended to this case, up to
the ``mode by mode'' re-scaling of the
$\beta_{JJ}$ functions and the corresponding canonical momenta.
Therefore, when analysing correlations and diagonalization of
the reduced density matrix within
the WKB branch corresponding to an expanding solution of the
Hamilton-Jacobi equation, (26), the results obtained in the previous
paragraphs
can be extended to encompass the longitudinal modes as well.
The conclusions concerning the conditions (\ref{adyab})--(\ref{Sdecoh}),
i.e.
(45)--(47), are thus valid in the
case of expanding closed FRW models with massive Abelian vector fields.
The same applies to (48)--(50).
However, for
contracting solutions of the Hamilton-Jacobi equation, there is a
problem due
to the $1 + 1/a$ factor. The equation corresponding to
(\ref{fakekg}) is the following
\beq
\ddot{\varphi}_{JJ} +  \left( \frac{-1 + f_{JJ}(a)}{\varphi_{JJ}}\right)
\dot{\varphi}_{JJ}^2 + \Omega^2_{JJ} \varphi_{JJ} = 0~.
\label{fakekg1aa}
\eeq
As it stands, we do not know of any exact or even adiabatic solution
of this equation, which is of the type
\beq
\varphi_{JJ}^{in} = \hat{\alpha}_{JJ} \varphi_{JJ}^{out}
+ \hat{\beta}_{JJ} (\varphi_{JJ}^{out})^*,
\label{bogo1aa}
\eeq
together with (\ref{outmodes}).
It is therefore difficult to draw any conclusions about correlations as
well
as diagonalization (and
localization) when the longitudinal modes are taken into consideration
in the
case of contracting WKB solutions.

This  problem becomes somewhat more acute
 when we try to analyse the decoherence between the two
WKB histories for $n=\pm 1$. Once again, we do not know how to obtain
 a form similar to (\ref{Inn}) for
${\cal I}_{nn'}(a,a)$ as that would
involve adiabatic solutions of the type (\ref{outmodes}) and that seems,
for the moment, difficult to obtain for the longitudinal modes in a
WKB contracting solution.

A possible alternative to deal with the longitudinal modes and construct
WKB solutions, may involve computing the influence functional
(and the functions $B$)  assuming another approach.
Namely, that the vector field mass may be  almost
negligible\footnote{This
hypothesis
should, however, be considered with some care in view of eqs. (46),(47)
and their
implications regarding the presence of a massive vector field as
providing a
scale relatively to the decoherence process and correlation analysis.}
and
therefore the presence
of longitudinal modes could be treated using a perturbative
scheme \cite{ggga}, assuming that
the massive transversal modes are dominant as they are anyway present
in the massless case.
In Ref. \cite{ggg}, it was considered that the interaction between the
system
and the environment is
such that there is decoherence between different WKB branches. This has
been shown for specific models in Refs. \cite{bb,cm,cm1} for $N=1$  and
$N > 1$
minisuperspace cases.
However, the use of massive Abelian vector field models raises the
problem that for {\em some} situations
that cannot be so easily achieved and demonstrated.
Nevertheless, the factor $1 + 1/a$ indicates that some divergences will
be present in the computation of the influence functional. One sould try
to
cure them by some type of renormalization
procedure, eg. via a cut-off, although its nature seems to indicate
that one has to go beyond quantum cosmological models arising
from Einstein theory plus matter, possibly considering an
effective model arising from
higher- derivative theories of gravity or even string theories.

\section{Decoherence and Back-Reaction Processes
in the Presence of Massive Vector Fields with  $SO(3)$ Global Symmetry}

\indent

In this section we extend the analysis of decoherence and correlations
to massive vector fields with non-Abelian SO(3) global symmetry.
The corresponding Wheeler-DeWitt equation
(\ref{12}),(\ref{14}) can be rewritten as:
\begin{eqnarray}
{\cal H}^{{\rm eff}} \Psi[a,\mbox{\boldmath $A$}_{\mu}]
&=& \left[ \frac{1}{ 2M_P^2}
\frac{\partial^2}{\partial ~a^2}
- M_P^2\left(  a^2 - {4\Lambda\over 9\pi }~a^4
\right) \right. \nonumber\\
& & - \frac{\partial ^2}{\partial \chi^2}
+ {\overline{\alpha}\over 3\pi} \left[ \chi^2  - {3\pi\over
2\overline{\alpha}}\right]^2
+ 4\pi a^2 m^2~\left[ 1 + \sqrt{{2\overline{\alpha}\over
3\pi}}\chi\right]^2
\nonumber\\
&& \left. - \sum_J f_{J(ab)}(a)\left[
\frac{\partial^2}{\partial (\beta^{J(ab)^2})}
 -
\Omega^2_{J(ab)}(a)~
\beta_{J}^{(ab)}~\beta_{J}^{(ab)} \right] \right]\Psi[a,\mbox{\boldmath
 $A$}_{\mu}] = 0 ,
\label{12aa}
\end{eqnarray}
where
\begin{equation}
f_{J(ab)}(a) = \left\{ \begin{array}{cll}
\overline{\alpha}\pi & {\rm if} & J-L=\pm 1 \\
\overline{\alpha}\pi + \left[(-1)^{4J} 48\pi^3 J(J+1)/( (2J+1) 4 m^2)
\right]~~
\left[ 1 + (1/a(t))\right] & {\rm if} & J=L
\end{array}  \right. ,
\label{12ba}
\end{equation}
and
\begin{equation}
\Omega^2_{J(ab)}(a) = \left\{ \begin{array}{cll}
\{(4m^2 a^2/3\pi) + (L+J+1)^2  + 4 ~\left[ 1 +
\sqrt{{2\overline{\alpha}\over
3\pi}}\chi\right]^2/3\pi   \} /
\overline{\alpha}\pi & {\rm if} & J-L=\pm 1 \\
\frac{(4m^2 a^2/3\pi) + (2J + 1)^2 +1 + 4\left[ 1 +
 \sqrt{{2\overline{\alpha}\over
3\pi}}\chi\right]^2/3\pi}{  \overline{\alpha}\pi
+ \left[(-1)^{4J} 48\pi^3 J(J+1)/( (2J+1) 4 m^2) \right]~~
\left[ 1 + (1/a(t))\right]}
 & {\rm if} & J=L
\end{array}  \right. ,
\label{12ca}
\end{equation}
and, furthermore, we have set the ambiguity factor to vanish, $P=0$.
Our minisuperspace is now two-dimensional, the scale factor
and function $\chi(\eta)$ (parametrizing the  homogeneous modes of the
non-Abelian massive vector fields) being the
classical observable degrees of freedom. The environment corresponds, as
before, to the inhomogeneous modes, i.e., the
$\beta_{J}^{(ab)}$-functions.

Actually, only a few particular multi-dimensional minisuperspace models
have been considered from the point of
view of decoherence and correlations between coordinates and momenta.
The Kantowski-Sachs model (N = 2) with a cosmological constant
and massive inhomogeneous conformally coupled
scalar field modes was studied in Ref. \cite{cm1}, the Bianchi type-I (N
= 3)
with mass{\em less}  inhomogeneous conformally coupled
scalar field modes was studied in Ref. \cite{ggga} and in Ref.
\cite{llo},
an N = 2 model has been analysed  where the two classical degrees of
freedom
were the scale factor and homogeneous mode of a
minimally massless scalar field, the environment being associated
with the inhomogeneous perturbations of {\em another}
minimally massless scalar field. The analysis of multi-dimensional
minisuperspace models is, in particular,
also relevant as it provides
a possible relation between the notion of decoherence between WKB
branches and
the decoherence between histories in the the so-called
Consistent Histories approach \cite{ff} in terms of space-time histories
(see section V. of Ref. \cite{ggga}). However, no
arguments have yet been put forward to relate the diagonalization of
the reduced density matrix  within a given WKB branch
to the notion of decoherence of different histories.
Nevertheless, as pointed out in Ref. \cite{ggg}, the functional
${\cal I}_{nn}(a,\chi,...;a',\chi^\prime,...)$ can be related somehow
to the notion of decoherence between histories for the cases where
$N>1$.
A particular solution of the Hamilton-Jacobi
equation generates a N-1 parameter family of trajectories,
but there will be only one classical
trajectory passing through each point of the minisuperspace
generated by {\em that} solution of the Hamilton-Jacobi equation.
In this sense, ${\cal I}_{nn}(a,\chi,...;a',\chi^\prime,...)$
will strenghten the suppression of interference between histories
belonging to a given WKB branch as it produces a more efficient
diagonalization of the reduced
density matrix.
However, calculations of Ref. \cite{llo}
have shown that a sucessfull decoherence between histories associated to
${\cal I}_{nn}$ has not been achieved for any of the models considered
so far,
possibly due to their simplicity \cite{ggg}.

As far as our case is concerned, we can see from eqs. (\ref{12aa})--
(\ref{12ca}) that the typical quadratic potential
of massive conformally coupled scalar field with
homogeneous modes is now replaced by the double-well quartic potential
${\overline{\alpha}\over 3\pi} \left[ \chi^2  - {3\pi\over
2\overline{\alpha}}\right]^2$. The remaks made above concerning the
$1 + 1/a$ factor still apply here. Notice, however, that one needs to
consider
the $(ab)$-$SO(3)$ group indexes.
It follows in particular, that eqs.(\ref{use})--(\ref{wigf}) (and
subsequent
ones)
remain valid
provided we include the $SO(3)$ group indexes
(eg. $\Pi_J \rightarrow \Pi_{J(ab)}$) and the
$\chi(\eta)$ function together with the scale factor.
Generally speaking, {\em all} our working hypothesis, considerations and
arguments presented in the last section can be extended to the
non-Abelian
case. The Hamilton-Jacobi equation is now
\beq
-\frac{1}{2}\left(\frac{ \partial S}{\partial a}\right)^{\ 2} + V(a)
+   \frac{1}{2}\left(\frac{ \partial S}{\partial \chi}\right)^{\ 2}
 + {\overline{\alpha}\over 3\pi} \left[ \chi^2  - {3\pi\over
2\overline{\alpha}}\right]^2
+ 4\pi a^2 m^2~\left[ 1 + \sqrt{{2\overline{\alpha}\over
3\pi}}\chi\right]^2 = 0~,
\label{HJ1}
\eeq where
$V(a) = - a^2 + {4\Lambda\over 9\pi }~a^4$.

As  our minisuperspace is now two-dimensional, the
Hamilton-Jacobi equation (\ref{HJ1}) is expected to have a one-parameter
family of solutions, each one generating a family of classical
trajectories in minisuperspace, each WKB branch interpreted
as describing a whole family of
classical trajectories, i.e. a set of different universes (and
not a single one as for the N = 1 case). From the
Hamilton-Jacobi, eq. (\ref{HJ1}), it follows that
the trajectories in our N = 2 minisuperspace
are far more complicated than those of the N = 1 case (Section 3).
Moreover, the WKB time is now
defined as
\beq
\frac{d}{d \eta} = G^{AB} \frac{ \partial S}{\partial q_A}
\frac{ \partial }{\partial q_B}~~,
\label{newwkbtime}
\eeq
with $A,B =1,2$, $q_1 = a, q_2 = \chi$ and $G_{AB} = {\rm diag}(-1,1)$.
We shall have as many $\eta$-affine parameters as different values of
the $(n)$-parameter. Hence, different values of $(n)$ will lead to
different
definitions of time for the Schr\"odinger equation (\ref{Schdg}).
This implies  that the influence functional
in (\ref{red}),(\ref{infl}) is actually a functional of two histories.
The state $\psi_{J(n)}\left( a,\chi,\left\{\beta_{J(ab)}\right\}\right)$
can be interpreted, not as being simply a function
of a point in minisuperspace, but instead as a function of the
whole history, which corresponds to the only
trajectory that belongs to the $(n)$-WKB branch and goes
through that particular point
($a,\chi$). Such description is fairly similar to the one of the
Feynman-Vernon influence functional \cite{fv}.

Let us now address the issues of correlations and decoherence within
each
WKB branch and compute
the relevant influence functional for our N = 2 minisuperspace model.
We briefly describe the main framework and consider, as in Section 3,
the transversal and longitudinal modes separately. We adopt
the terminology of Ref. \cite{ggga} here as well. Defining new variables
$q^{1,2}_{A} = x_A \pm \frac{1}{2}y_A$ and {\em assuming} the Gaussian
Ansatz (\ref{gauss}) for each $\beta_J^{(ab)}$-mode, we can
write (omiting the $(ab)$ indexes for convenience)
\beq
{\cal I}_{(n,n)J}(q^1_A,q^2_A) =
\exp\left[-\epsilon_{(J)}^{AB}y_A y_B\right] \exp
 \left[i\tilde{\epsilon}_{(J)}^A y_A
\right],
\label{inn41}
\eeq
where $\epsilon^{AB}$ and $\tilde{\epsilon}^{A}$ are designated as
decoherence
matrix and phase vector respectively, and
\begin{eqnarray}
\epsilon_{(J)}^{AB} & = &  \frac{1}{4B_{rJ}}
[(B_{rJ})^{\prime A}(B_{rJ})^{\prime B }  + (B_{iJ})^{ \prime
A}(B_{iJ})^{\prime
 B }]~,
            \label{eps1}    \\
\tilde{\epsilon}_{(J)}^A & = &  (D_J)^{\prime A} - (B_{iJ})^{\prime A}/4
 (B_{rJ})^{\prime A}~.
 \label{eps2}
\end{eqnarray}

Let us again consider the minisuperspace with a region, say for large
values of $a$, for which an adiabatic solution of eq. (\ref{fakekg})
of the type (\ref{outmodes}) and (\ref{bogo}) can be established.
The quantum state of the environment
is determined by choosing the Bogolubov coefficients.
We then obtain \cite{ggga}
\begin{eqnarray}
\epsilon_{(J)}^{AB} & = &  \Omega_{(J)}^{\prime A} \Omega_{(J)}^{\prime
B } +
\eta^{\prime A} \eta^{\prime B }  \Omega^2_J |\hat{\beta}_J|^2~,
            \label{eps1a}    \\
\tilde{\epsilon}_{(J)}^{A} & = &  \eta^{\prime
A}\left[\frac{\Omega_J}{2} +
 |\hat{\beta}_J|^2\right]
+  \frac{1}{2\Omega_J} \left[\frac{
\dot{\Omega_J}}{4\Omega_J}\right]^{\prime
 A}.
\label{eps2a}
\end{eqnarray}
The first term in (\ref{eps1a}) corresponds to the adiabatic vacuum
contribution ($\hat{\beta}_J=0$) while the second one is related to
particle
 creation.
It is interesting to notice  that for N = 2 one can always\footnote{For
$N>2$
the situation is however, different because
$\epsilon^{AB}$ is positive definite and also degenerate whenever $N>2$
as the directions of the $(B_{rJ})^{\prime A}$ and $(B_{iJ})^{\prime A}$
vectors
diagonalize the reduced density matrix and therefore any other
orthogonal direction to these is an eigenvector with null eigenvalues.}
generate a diagonalization along two independent directions of
minisuperspace by coupling to a variable mass.

Correlations between each minisuperspace coordinate $a$ and $\chi$ and
their canonical momenta
can be analysed by examining peaks in the {\em reduced} Wigner function
\beq
F_{W1} (q^A,\mbox{\boldmath $\pi$}_{q_{1}}) =
\int d\mbox{\boldmath $\pi$}_{q_{2}} F_{W(n)}(q^A; \mbox{\boldmath
 $\pi$}_{q_{A}})
\label{redwigf}
\eeq
which is equivalent to
\beq
F_{W1} (q^A,\mbox{\boldmath $\pi$}_{q_{1}}) =
\exp \left[\epsilon^{11}\left[ \mbox{\boldmath $\pi$}_{q_{1}} - M_P^2
 \frac{\partial S}{\partial q^1} -
\tilde{\epsilon}^1\right]^2\right],
\label{redwigfa}
\eeq
where $\mbox{\boldmath $\pi$}_{q_{1}}$ is the momentum conjugate to
$q^A,A=1$.
The strong correlation condition translates as
\beq
\epsilon^{AA} \ll \overline{\mbox{\boldmath $\pi$}_A}^2
\eeq
(where $\overline{\pi_A}$ is a typical value of the momentum along
the trajectory) and the
strong decoherence condition is
\beq
\epsilon^{AA} \gg 1/q^A.
\eeq
Notice that eqs. (\ref{redwigf}),
(\ref{redwigfa}) must include the summation over the $SO(3)$-group
indexes and
{\em all} modes, using eqs. (\ref{inn41}) to (\ref{eps2a}). The same
holds
for eqs. (\ref{Inn})--(\ref{Inn2}) when analysing decoherence between
different WKB branches.

The above construction is valid  for the transversal modes
as those behave similarly to massive conformally
coupled scalar fields. Let us take the
case for which $\hat{\beta}_J=0$, i.e., we choose the quantum
state of the environment modes to be the adiabatic vacuum. In this case
we must put $\Lambda = 0$. The relevant quantities to analyse the
decoherence
and correlation will be
$\epsilon^{11}_J$ and $\epsilon^{22}_J$
\beq
\epsilon^{11}_J \simeq \frac{\left(\frac{4m^2}{3\pi}\right)^2
 a^2}{\frac{4m^2a^2}{3\pi}
+ (L+J+1)^2  +  4 ~\left[ 1 + \sqrt{{2\overline{\alpha}\over
3\pi}}\chi\right]^2}~,
\eeq
\beq
\epsilon^{22}_J \simeq \frac{{2\overline{\alpha}\over
3\pi}\left[ 1 + \sqrt{{2\overline{\alpha}\over
3\pi}}\chi\right]^2}{\frac{4m^2a^2}{3\pi}
+ (L+J+1)^2  +  4 ~\left[ 1 + \sqrt{{2\overline{\alpha}\over
3\pi}}\chi\right]^2}~.
\eeq
The sum in $J$ implies that, similarly to Ref. \cite{ggg},
$\epsilon^{11}$ and
$\epsilon^{22}$ behave, for large $a$, proportionally to $a$ and
$a^{-1}$,
respectivelly. Notice that one expects the $\chi$-field to evolve
towards
one of the
minima of the potential ${\overline{\alpha}\over 3\pi} \left[ \chi^2  -
 {3\pi\over
2\overline{\alpha}}\right]^2$ for large values of $a$, following Ref.
[16], and
neglecting curvature terms as $a \rightarrow \infty$.
 From the Hamilton-Jacobi equation we obtain that a typical
value along a WKB trajectory for the canonical conjugate momenta
to $a$ and $\chi$ are
$\mbox{\boldmath $\pi$}_{a} = \frac{\partial S}{\partial a} \sim a$
and
$\mbox{\boldmath $\pi$}_{\chi} = \frac{\partial S}{\partial \chi} \sim
\chi^2$,
assuming a small mass in order to neglect the interaction terms.
Such approach has already been discussed in
Section 3 regarding the longitudinal modes.
As far as WKB expanding solutions are concerned, conditions (65) and
(66)
with (67) and (68) seem to indicate that, for the $\chi$-field, the
strong
correlation condition
will be satisfied but not the strong decoherence one. For the scale
factor,
the strong correlation and strong decoherence conditions will be
satisfied
in the very sense of Ref. \cite{ggga}. We shall discuss the apparent
failure in fulfilling the strong decoherence condition for the
$\chi$-field
in Section 5.

Once again, the longitudinal modes are well behaved concerning
diagonalization and correlation within a suitable  expanding WKB branch
and the
above results are equally suited here. However, the
difficulty associated to contracting branches still remains.

Equations (48)--(50), including the $(ab)-SO(3)$ indexes, are valid for
the
 transversal modes.
The same conclusion holds for longitudinal modes within expanding WKB
branches.
If we consider two different WKB expanding branches in our N-2
dimensional
 minisuperspace
with an adiabatic vacuum, eq. (48) can be reduced to
\beq
{\cal I}_{nn'}(a,a)
 \simeq  \exp \left\{ \sum_{J(ab)} \left[
 -
\frac{1}{64} \frac{(\nabla S_{(n)} -  \nabla S_{(n')})(\nabla
 \Omega_{J}^{(ab)})^2}{\Omega^{(ab)4}_{J}}
\right]\right\},
\label{Inn111}
\eeq
with $\nabla S_{(n)} \nabla \equiv G^{AB} \frac{ \partial S}{\partial
q_A}
\frac{ \partial }{\partial q_B}$. Hence, we find that
the interference between terms with different $(n)$ is
exponentially supressed,
depending on how different is the WKB time variation of
$\Omega_{J}^{(ab)}$
 along the
two trajectories (cf. comment after eq. (57)). However, when there are
two branches and one corresponds
to an expanding solutions while the other to contractiong ones,
the issues raised
in Section 3 with respect to longitudinal modes equally will apply in
this case
 as well.
Namely, one cannot use (48)--(50) unless some consistent
perturbative scheme to treat the longitudinal modes
when $m\ll 1$ is available and justifiable.

 \section{Conclusions and Discussion}

\indent

We have discussed the quantum cosmology of a massive vector field
coupled with gravity and we have shown that the resulting model
possesses interesting properties in what concerns the decoherence of the
scale factor of a closed FRW geometry. In the presence of a massive
vector
field with U(1) global symmetry, the scale factor is the only decohered
quantity while for the non-Abelian case with SO(3) global symmetry,
the scale factor and the homogeneous mode $\chi(t)$ are the decohered
variables expected to behave classically.

As far as we consider expanding semiclassical solutions,
 the models we propose
can be regarded, to a certain extent, on the same footing
as the ones where decoherence of degrees of freedom of
the metric is achieved via tracing out higher modes of self-interacting
scalar
fields. The inhomogeneous modes of massive vector fields, which
were expanded in spin-1 hyperspherical harmonics,
represent an interesting alternative to play the role
of environment for the metric in the
Abelian case and for
the metric and the homogeneous mode of the non-Abelian massive vector
field.
However, in the latter case, we find that the strong decoherence
condition,
eq. (66), for the $\chi$-field (parametrizing the non-Abelian
massive vector field homogeneous modes) is not satisfied.
Unfortunately, since the literature on $N>1$ minisuperspace models
is rather scarce, we could not contrast our results with the existing
ones.
We can mention nevertheless, that, for instance, in Ref. [8],
in a diagonal
Bianchi type-I model where the environment consists of modes of a
massless
conformally coupled scalar field,
a problem of similar nature to ours is encountered. One could speculate
whether
the decoherence and correlation conditions would be satisfied using
solutions
that already account for the back-reaction, rather than the classical
histories
or, instead, considering higher-derivative terms as done in
Ref. \cite{ggga}. A self-consistent approach along these lines
could be seen as a way to implement our massive
non-Abelian vector field model.

We have also
shown that, for the transversal modes, known techniques and the
associated discussion on decoherence and localization within WKB
branches,
namely using the Gaussian Ansatz for the wave function and computing the
Wigner functional in the adiabatic limit, still
holds for both cases we have considered (Abelian and non-Abelian). The
same
can be said for the longitudinal modes at least in what concerns
expanding models. Issues related with the problem of treating the
longitudinal modes in contracting models, in particular regarding
the decoherence between different WKB branches,  have been discussed and
some possible fixes have been sugested.

\bigskip
\bigskip
\bigskip

{\large\bf Acknowledgments}

\bigskip
\indent
The authors gratefully acknowledge H.F. Dowker, B.L. Hu and
J.P. Paz for enlightening discussions, J.S. Dowker for correspondence,
U. Ancarani, L. Garay and C. Kiefer for conversations on issues related
to this
paper, J. Mour\~ao for valuable comments and M.C. Bento for suggestions.

\vfill\eject

\pagebreak
\noindent
{\bf Appendix}

\indent

The full effective Hamiltonian density is given by:

\begin{eqnarray*}
{\cal H}^{\rm {eff}} & = & N \Bigg\{ -\frac{1}{2} \mbox{\boldmath $\pi$}^2_a -
a^2
+ \frac{4\Lambda}{9\pi M^2_P} a^4 + 4\pi a^2
\frac{m^2}{M^2_P} \left[ 1 + \sqrt{\frac{2\bar \alpha}{9\pi}}\chi
\right]^2  + \frac{4}{3\pi} \frac{m^2}{M^2_P}
 \beta^{bcNM}_{LJ} \beta^{bcLJ}_{NM} \nonumber \\ \nonumber \\
&& + \frac{1}{2\pi \bar \alpha} \left[
 1 + \sqrt{\frac{2\bar \alpha}{3\pi}} \chi \right]
\left[ \beta^{da NM}_{LJ}
\beta^{bd N^\prime M^\prime}_{L^\prime J^\prime}
 + \beta^{db NM}_{LJ}
\beta^{da N^\prime M^\prime}_{L^\prime J^\prime} \right]
\sigma^m_a \sigma^{m^\prime}_{b}
 S_5\left[
\matrix{
NM & N^\prime M^\prime & m \cr
LJ & L^\prime J^\prime & m^\prime}\right] \nonumber \\ \nonumber \\
&& + \frac{1}{2\pi\bar\alpha}
\left[ 1 + \sqrt{\frac{2\pi}{3\pi}} \chi \right]^2
\left[ \beta^{da NM}_{LJ}
\beta^{bd N^\prime M^\prime}_{L^\prime J^\prime} +
\beta^{db NM}_{LJ} \beta^{da N^\prime M^\prime}_{L^\prime J^\prime}
\right]
\sigma^m_a \sigma^{m^\prime}_b S_5 \left[
\matrix{
NM & N^\prime M^\prime & m \cr
LJ & L^\prime J^\prime & m^\prime}\right]  \nonumber \\ \nonumber \\
&& - \frac{1}{\pi \bar \alpha} ~
\beta^{ce NM}_{LJ}~
\beta^{dc N^\prime M^\prime}_{L^\prime J^\prime}~
\beta^{ed N'' M''}_{L'' J''}~
\varepsilon_{dab} \sigma^m_d \sigma^{m^\prime}_a \sigma^{m''}_b
S_6  \left[ \matrix{
LJ & L^\prime J^\prime & L''J'' & m \cr
MN & M^\prime N^\prime & N'' M''& m^\prime m''} \right]  \nonumber \\
\nonumber \\
&& + \frac{2}{\pi \bar \alpha}
\left( \beta^{de NM}_{LJ}
\beta^{fe N^\prime M^\prime}_{L^\prime J^\prime}
\beta^{cf N'' M''}_{L'' J''} -
\beta^{de NM}_{LJ}
\beta^{cf N^\prime M^\prime}_{L^\prime J^\prime}
\beta^{fe N''M''}_{L''J''} \right)
\left[ 1 + \sqrt{\frac{2 \bar \alpha}{3\pi}} \chi \right]  \times
\nonumber \\
&& \times \varepsilon_{dac} \sigma^{m^\prime}_a
S_6  \left[ \matrix{
LJ & L^\prime J^\prime & L''J'' & m \cr
MN & M^\prime N^\prime & N'' M''& m^\prime m''} \right]  \nonumber \\
\nonumber \\
&& + \frac{2}{\pi \sqrt{\bar \alpha}}
\beta^{be NM}_{LJ}
\beta^{de  N^\prime M^\prime}_{L^\prime J^\prime}
\left[ 1 + \sqrt{\frac{2 \bar \alpha}{3\pi}} \chi^2 \right]
\varepsilon_{dab} S_7
\left[  \matrix{
LJ & L^\prime J^\prime & m \cr
NM & N^\prime M^\prime & m^\prime} \right]_a + \nonumber \\ \nonumber \\
&& +\frac{2}{\pi
\bar \alpha}
\beta^{ce NM}_{LJ}
\beta^{dc N^\prime M^\prime}_{L^\prime J^\prime}
\beta^{ed N''M''}_{L''J''} \sigma^{m^\prime}_a~
S_8 \left[
\matrix{
LJ & L^\prime J^\prime & L''J'' & m\cr
NM & N^\prime M^\prime & N'' M''& m m^\prime} \right]_a
\nonumber \\ \nonumber \\
&&+ \frac{1}{\pi \bar\alpha}\left( \beta^{dc LJ}_{NM}
\beta^{ed \bar L \bar J}_{\bar N \bar M}
\beta^{fc L^\prime J^\prime}_{N^\prime M^\prime}
\beta^{ef \bar L^\prime \bar J^\prime}_{\bar{N}' \bar{ M}'} -
\beta^{dc LJ}_{NM}
\beta^{ed \bar L \bar J}_{\bar N \bar M}
\beta^{ef \bar L^\prime \bar J^\prime}_{N^\prime M^\prime}
\beta^{fc \bar L^\prime \bar J^\prime}_{\bar N' \bar M'}
\right) \times \nonumber \\
&&\times S_9
\left[
 \matrix{
LJ & \bar L \bar J & m &L^\prime J^\prime & \bar L^\prime \bar J^\prime
& m\cr
NM & \bar N \bar M &m & N^\prime M^\prime &\bar N^\prime \bar M^\prime &
m^\prime} \right]  \nonumber \\ \nonumber \\
&&+ \mbox{\boldmath $\pi$}^2_\chi + \frac{\bar \alpha}{3 \pi} \left[ \chi^2 -
\frac{3\pi}{2\bar \alpha}
\right]^2 + \frac{1}{\pi \bar \alpha}~
\beta^{bc NM}_{LJ}  \beta^{bc LJ}_{NM} (L + J + 1)^2  \nonumber \\
\nonumber \\
&&+ \frac{4}{\pi \bar\alpha} \left[ 1 + \sqrt{\frac{2 \bar
\alpha}{2\pi}} \chi
\right]^2 \beta^{de NM}_{LJ} \beta^{de LJ}_{NM} +
\pi \bar \alpha \left( \mbox{\boldmath $\pi$}_{ \beta^{bc LJ}_{NM}}
\mbox{\boldmath $\pi$}_{\beta^{bc NM}_{LJ}} + \mbox{\boldmath $\pi$}_{\beta^{bc
JJ}_{ NM}}
\mbox{\boldmath $\pi$}_{\beta^{bc JJ}_{NM}}  \right) \Bigg\} +\nonumber \\
\nonumber \\
&& + \frac{\sqrt 2}{\pi \sqrt{\pi\bar \alpha}}
a~\alpha^{cd J^\prime N^\prime}_{M^\prime} \beta^{db NM}_{JJ}~
\mbox{\boldmath $\pi$}_\chi \varepsilon_{bac} \sigma^m_a S_1 \left[ \matrix{JJ
& M^\prime &
N^\prime
\cr NM & J^\prime & m} \right] \nonumber \\ \nonumber \\
&& + 2a \left[ 1 +
\sqrt{\frac{2\bar \alpha}{3\pi}} \chi \right]
\alpha^{dc J^\prime N^\prime}_{M^\prime} \mbox{\boldmath $\pi$}_{\beta^{bc
JJ}_{MN}}
\varepsilon_{bad} \sigma^m_a S_1 \left[ \matrix{ JJ &
M^\prime N^\prime \cr NM& J^\prime m} \right]  \nonumber \\ \nonumber
\\
&&+ 4a~\alpha^{cd J''N''}_{M''} \beta^{db L^\prime J^\prime}_{M^\prime
N^\prime} \mbox{\boldmath $\pi$}_{\beta^{db LJ}_{NM}} S_2 \left[ \matrix{NM &
N^\prime
M^\prime & J'' & m \cr  LJ& L^\prime J^\prime & M''N'' & m } \right]
\nonumber \\
\nonumber \\
 && + (-1)^{2J} a~\alpha^{bc JN}_{M} \mbox{\boldmath $\pi$}_{\beta^{bc
JJ}_{NM}}
\sqrt{\frac{J(J+1)16\pi^2}{2J+1}}\nonumber \\ \nonumber \\
&& - \frac{4}{3\pi}\frac{m^2}{M^2_P} a^4 \frac{\alpha^{bc J M}_{N}~
\alpha^{bc J N}_{M}}{N}  \nonumber \\ \nonumber \\
&& +\frac{4a^2}{\pi\bar\alpha}~
\frac{\alpha^{cd \bar J \bar N}_{\bar M}  \alpha^{cd'J'N'}
_{M^\prime}}{N}
\beta^{db N^\prime M^\prime}_{L^\prime J^\prime}
\beta^{d^\prime b \hat N^\prime \hat M^\prime}_{\hat L^\prime \hat
J^\prime}~S_2
\left[ \matrix{NM & N^\prime M^\prime & \bar J& m \cr LJ & L^\prime
J^\prime & \bar
M \bar N & m} \right]\times \nonumber \\
&&\times
S_2 \left[ \matrix{ LJ & \hat N^\prime \hat M^\prime & \bar J ^\prime &
m^\prime
\cr
NM & \hat L^\prime \hat J^\prime & \bar M^\prime \bar N^\prime & m^\prime}
\right]
\nonumber \\ \nonumber \\
&& - \frac{4 a^2}{\pi \bar \alpha}
\frac{\alpha^{cd \bar J \bar N}_{ \bar M} \alpha^{cd \bar J^\prime \bar
N^\prime}
_{\bar M^\prime}}{N} \beta^{db N^\prime M^\prime}_{J^\prime J^\prime}
\beta^{d^\prime b \hat N^\prime \hat M^\prime}_{\hat J^\prime \hat J}
S_2 \left[
\matrix{ NM & N^\prime M^\prime &\bar J & m \cr JJ & L^\prime J^\prime &
\bar M \bar
N & m}\right]\times \nonumber \\
&&\times S_2 \left[ \matrix{JJ & \tilde N^\prime \tilde M^\prime
& \bar J^\prime
& m \cr NM & \hat J \hat J^\prime & \bar M^\prime \bar N^\prime & m }
\right]
\nonumber \\ \nonumber \\
&&+  \frac{4a^2}{\pi \bar \alpha}~\frac{\alpha^{cd \bar J
\bar N}_{\bar M} \alpha^{d'c\bar J \bar M}_{\bar M}}{N}
\left[ 1 + \sqrt{\frac{2 \bar \alpha}{3\pi}} \chi \right] \beta^{db
\tilde
J^\prime \tilde L^\prime}_{\tilde N^\prime \tilde M^\prime}
\varepsilon_{bad^\prime}
\sigma^m_a~S_1 \left[ \matrix{ \bar J \bar N & J & L \cr m \bar M & N & M}
\right]\times \nonumber \\
&&\times
S_2\left[ \matrix {NM & \tilde M^\prime \tilde M^\prime & \bar M^\prime &
m^\prime \cr
JJ & \tilde J^\prime \tilde L^\prime & \bar J^\prime N & m^\prime} \right]
\nonumber \\ \nonumber \\
&& - \frac{4a^2}{\pi \bar \alpha}~~\frac{\alpha^{cd \bar J^\prime
N^\prime}_{M^\prime}~\alpha^{d^\prime c \bar J \bar N}_{\bar M}}{N}
\left[ 1 + \sqrt{\frac{2\bar \alpha}{3\pi}} \chi \right] \beta^{bd
\tilde J \tilde
L^\prime}_{\tilde N^\prime \tilde M ^\prime}~\varepsilon_{bad^\prime}
\sigma^m_a~S_1 \left[ \matrix{ \bar J \bar N & J & J \cr m \bar M & N & M
}\right]\times \nonumber \\
&&\times
S_2\left[ \matrix {NM & \tilde M^\prime \tilde M^\prime & \bar M^\prime &
m\prime \cr
JJ & \tilde J^\prime \tilde L^\prime & \bar J^\prime N & m^\prime} \right]
\nonumber
\\ \nonumber \\
&& - \frac{a^2}{\pi \bar \alpha} ~ \frac{\alpha^{dcyN}_M~\alpha^{bc\bar
J \bar
N}_{\bar M}}{N} \left[ 1 + \sqrt{\frac{2\bar \alpha}{3\pi}} \chi \right]
(-1)^{2
\bar J}~\sqrt{\frac{\bar J(\bar J+ 1) 16 \pi^2}{2\bar J +
1}}~\varepsilon_{bad}
\sigma^m_a~S_1 \left[ \matrix{JJ & \bar J & m \cr NM & \bar M & \bar N}
\right]
\nonumber \\ \nonumber \\
&&- \frac{4a^2}{\pi \bar \alpha} ~ \frac{\alpha^{dc\bar J \bar N}_{\bar
M}~\alpha^{d^\prime c \bar J^\prime \bar N^\prime}_{\bar
M^\prime}}{N}~\left[ 1 +
\sqrt{\frac{2\bar \alpha}{3\pi}} \chi \right]^2 \varepsilon_{ba^\prime
d}
\sigma^{m^\prime}_{a^\prime}~S_1 \left[ \matrix{JJ & \bar J & \bar N \cr
NM & \bar M
& m} \right]~S_1 \left[ \matrix{ NM & \bar J^\prime & \bar N^\prime \cr JJ
& \bar
M^\prime & m^\prime} \right] \nonumber \\ \nonumber \\
&& - \frac{4a^2}{\pi \bar \alpha}~ \frac{\alpha^{cd \tilde J \tilde
N}_{\tilde
M}~\alpha^{cd^\prime \bar J \bar N}_{\bar M}}{N}~\beta^{db \tilde N
\tilde
M^\prime}_{\hat L^\prime \hat J^\prime}~\beta^{d^\prime b N^\prime
M^\prime}_{L^\prime J^\prime}~S_2 \left[ \matrix{ NM & \hat N^\prime
\hat M^\prime
& \tilde J \tilde P & m \cr JJ \hat L^\prime \hat J^\prime & \tilde M
&\tilde M &
m} \right] \times \nonumber \\
&&\times S_2 \left[ \matrix{ NM & N^\prime M^\prime & \bar
J \bar N & m^\prime \cr JJ & L^\prime J^\prime & \bar M & \bar m}
\right] \nonumber
\\ \nonumber \\
&& -\frac{2a^2}{\pi \bar \alpha}~ \frac{\alpha^{bc J N}_M \alpha^{cd
\bar J \bar
N}_{\bar M}}{N} \beta^{db N^\prime M^\prime}_{L^\prime J^\prime} (-1)^{2
\bar J}
\sqrt{\frac{\bar J(\bar J + 1) 16\pi^2}{2 \bar J + 1}}S_2 \left[
\matrix{ NM &
N^\prime M^\prime & \bar J \bar N & m \cr JJ & L^\prime J^\prime & \bar
M & m}
\right] \nonumber \\ \nonumber \\
&&+ \frac{a^2}{\pi \bar \alpha} \left[ \frac{\alpha^{dc \bar J \bar
N}_{\bar M}
\alpha^{cf \tilde J \tilde N}_{\tilde M}}{N} \beta^{fb NM}_{LJ} -
\frac{\alpha^{dc
\bar J \bar N}_{\bar M} \alpha^{bf \tilde J \tilde N}_{\tilde M}}{N}
\beta^{fc
NM}_{LJ} \right] \left[ 1 + \sqrt{\frac{2 \bar \alpha}{3\pi}} \chi
\right]
\varepsilon_{bad}~ \sigma^m_a~\times \nonumber \\
&&\times
S_3 \left[ \matrix{\bar J m & NM & \tilde J
\tilde N
\cr \bar M \bar N & LJ & \tilde M} \right] \nonumber \\ \nonumber \\
&& + \frac{a^2}{\pi \bar \alpha} \left[ \frac{\alpha^{df \bar J \bar
N}_{\bar M}
\alpha^{dg \tilde J^\prime \tilde N^\prime}_{\tilde M^\prime}}{N}
\beta^{fe NM}_{LJ}
\beta^{gc \tilde N \tilde M}_{\tilde L \tilde J} -
\frac{\alpha^{df \bar J \bar N}_{\bar M} \alpha^{gc \tilde J^\prime
\tilde N'}_{\tilde M}}{N} \beta^{fcNM}_{LJ} \beta^{dg \tilde N \tilde
M}_{\tilde L
\tilde J}~\right] \times \nonumber \\
&&\times S_4 \left[ \matrix{ NM & \bar Jm & \tilde N \tilde
M & \tilde J^\prime \tilde J \cr LJ & M \bar N & m \tilde L & \tilde
M^\prime \tilde
N^\prime} \right], \nonumber \\
 \end{eqnarray*}
where
\begin{eqnarray*}
&& S_1 \left[ \matrix {LJ & M^\prime & m \cr NM & J^\prime & N^\prime}
\right] =
\int d^3x \sqrt{^{s^3} g}
Y^{1NJ}_{mNM}
D^{J^\prime\phantom{M^\prime}N^\prime}_{\phantom{J^\prime}M^\prime},
\nonumber \\
\nonumber \\
&& S_2\left[ \matrix{ NM & N^\prime M^\prime & J''N'' & m \cr LJ &
L^\prime J^\prime & M'' 1 & m} \right] = \int d^3x \sqrt{^{s^3}g}
Y^{1NM}_{mLJ}
Y^{mN^\prime M^\prime}_{1 L^\prime J^\prime} D^{J''
\phantom{M''}N''}_{\phantom{J''}M''},
\nonumber \\ \nonumber \\
&& S_3 \left[ \matrix{ NJ & J^\prime N^\prime & M'' \cr LJ & M^\prime
J'' & N''}
\right] = \int d^3x \sqrt{^{s^3}g} Y^{1NM}_{mLJ}
D^{J^\prime\phantom{M^\prime}N^\prime}_{\phantom{J^\prime} M^\prime}
D^{J''\phantom{M''}N''}_{\phantom{J''}M''}, \hfill \cr \cr
&& S_4 \left[ \matrix{NM & J^\prime N^\prime & L''J'' & J'''N''' \cr LJ
& M^\prime
m & M''N''& M''' m }\right] = \int d^3x \sqrt{^{s^3 }g}
Y^{1NM}_{mLJ} D^{J^\prime \phantom{M}N^\prime}_{\phantom{J^\prime} M}
Y^{mL'' J''}_{1 N'' M''}
D^{J'''\phantom{M'''}N'''}_{\phantom{J'''}M'''},
\nonumber \\ \nonumber \\
&&S_5 \left[ \matrix{NM & N^\prime M^\prime & m \cr LJ & L^\prime
J^\prime &
m^\prime} \right]
= \int d^3x \sqrt{^{s^3}g} Y^{1 NM}_{m LJ}~Y^{1 N^\prime
M^\prime}_{m^\prime~L^\prime J^\prime}, \nonumber \\ \nonumber \\
&& S_6 \left[ \matrix {LJ& L^\prime J^\prime & L'' J'' & m \cr
NM & M^\prime N^\prime & M'' N''& m^\prime m''} \right] =
\int d^3x \sqrt{^{s^3}g} Y^{1 L J}_{m N M}
Y^{1 L^\prime J^\prime}_{m^\prime M^\prime N^\prime}
Y^{1 L'' J''}_{m^\prime M'' N''}, \nonumber \\ \nonumber \\
&& S_7 \left[ \matrix{ LJ & L^\prime J^\prime & m \cr MN & M^\prime
N^\prime &
m^\prime }\right]_a = \int d^3 x \sqrt{^{s^3}g} \frac{\partial
Y^{1LJ}_{mNM}}{\partial x^i}~L^i_a Y^{m^\prime L^\prime J^\prime}_{1
M^\prime
N^\prime}, \nonumber \\ \nonumber \\
&& S_8 \left[ \matrix {LJ & L^\prime J^\prime & L''J''&m \cr MN &
M^\prime N^\prime
& M''N'' & m~m^\prime} \right]_a =
\int d^3x~ \sqrt{^{s^3}g}~ \frac{\partial Y^{1LJ}_{mNM}}{\partial
x^i}~L^i_a~Y^{1L^\prime J^\prime}_{m^\prime N^\prime}~Y^{m L'' J''}_{1
N'' M''},
\nonumber \\ \nonumber \\
&& S_9 \left[ \matrix{LJ&\bar L \bar J & L^\prime J^\prime & \bar
L^\prime \bar
M^\prime & m \bar m \cr
NM & NM & N^\prime M^\prime & \bar N^\prime \bar M^\prime & m \bar m}
\right]
= \int d^3 x \sqrt{^{s^3}g}~Y^{1LJ}_{mNM}~J^{\bar m \bar L \bar J}_{1
\bar M \bar
N}~Y^{m L^\prime J^\prime}_{1 M^\prime N^\prime}~Y^{1 \bar L \bar
J^\prime}_{\bar
m \bar M^\prime \bar N^\prime},
\nonumber \\
 \end{eqnarray*}
and
$$
Y^{1NM}_{mLJ} = \sqrt{\frac{(2L+1)(2J+1)}{16 \pi^2}}D^{L
\phantom{N}N^\prime}_{\phantom{L} N} \left( \matrix{ L & J & j \cr
N^\prime & M &
m}\right)~,
$$
where $D^{L \phantom{N} N^\prime}_{\phantom{L} N}$ is a representation
for the
scalar  harmonics
 $$
Q^n_{\ell m} = \pi^n_\ell(\chi) Y_{\ell m}(\theta, \phi)~,
$$
$$
\pi^m_\ell (\chi) = \sin^\ell \chi \frac{d^{\ell + 1} (\cos n
\chi)}{d(\cos
\chi)^{\ell + 1}}
$$
are Fock harmonics, $Y_{\ell m}(\theta,\phi)$ are spherical
harmonics on $S^2$
and $\left( \matrix{L&J&J \cr N^\prime & M & m}\right)$ are $3-j$
symbols. $\sqrt{^{s^3}g}$ denotes the square root of the determinant of the
metric over the unitary 3-sphere and $L^i_a$ represents the transformation
between
the left-invariant basis on $S^3$ and a coordinate basis.
 We have
used
$$
\int d^3x~\sqrt{s^3 g} Y^{1NM}_{mLJ}~Y^{m N^\prime M^\prime}_{1 L^\prime
J^\prime} = \delta^N_M \delta^M_{M^\prime} \delta^J_{J^\prime}
\delta^L_{L^\prime}
$$
and
$$
\sigma^a_mL^i_a \partial_i D^{J \phantom{N} M}_{\phantom{J}N} =
(-1)^{2J}
\sqrt{\frac{J(J+1)16\pi^2}{2J+1)}} Y^{1JM}_{mnJ}~,
$$
and also that
$$
{^{s^3}\!\!\!g} = ^{s^3}\!\!\!g_{ab} \omega^a \otimes \omega^b = c_{mn}
\omega^m_p
\otimes
\omega^n_p
$$
with
$$
c_{mn} = \left( \matrix{ 0 & 0 & 1 \cr
0 & -1 & 0 \cr
1 & 0 & 0} \right)~.
$$
Furthermore, we have made use of the relation provided by the equation of
motion:
$$
\left( D^iF^{\phantom{i0}(bc)}_{i0} + m^2 A^{\phantom{0}(bc)}_{0}
\right) T_{bc} =
0~$$
i.e.,
$$\left( \partial^i F^{\phantom{i0}(bc)}_{i0} + m^2
A_0^{\phantom{0}(bc)}\right) +
\left\langle \left[A^i, F_{i0} \right] \right\rangle_{(bc)} = 0,
$$
where $\langle ~~ \rangle_{(bc)}$ means the $(bc)$-component projection.
Using
the expansion of $ \mbox{\boldmath $A$}= A^{(bc)}_\mu \omega^\mu T_{bc}$ in the
above
equation,
multiplying by $D^{J \phantom{M}N}_{\phantom{J}M}$ and integrating
over $S^3$ we get after integration by parts (cf. ref. [36]):
\begin{eqnarray*}
\alpha^{bcM}_{JN} &+& \frac{(-1)^{2J}}{\sigma^2a}
\sqrt{\frac{J(J+1)16\pi^2}{2J+1}}~\frac{2N\pi \bar
\alpha}{m^2}~\pi_{\beta^{bcNM}_{JJ}} \nonumber \\
& + & \int_{s^3} \frac{d^3x \sqrt{^{s^3}g}}{m^2} \left\langle \left[A^i,
F_{0i}
\right] \right\rangle_{(bc)} \cdot D^{J \phantom{M}N}_{\phantom{J}M} =
0~.
\end{eqnarray*}
 From the last term of the last equation we get integrals of the type $S_1,
S_2, S_3,
S_4$;  the first two terms are valid only for the Abelian case (where
the
$(ab)$-SO(3) group indices have been obviously disregarded).


\begin{thebibliography}{99}
\bibitem{aaa} A.O. Caldeira and A.J. Legget, {\it Physica} {\bf A121}
(1983)
587; {\it Ann.Phys.} (N.Y.) {\bf 149} (1983) 374; \\
E. Joos and H.D. Zeh, \ZP {\bf B59} (1985) 223; \\
W. Unruh and W.H. Zurek, \PR {\bf D40} (1989) 1071; \\
W.H. Zurek, \PR {\bf D24} (1981) 1516, {\em ibid} {\bf D26} (1982) 1862.

\bibitem{bb} J.J. Halliwell, \PR {\bf D36} (1987) 3626.

\bibitem{cc} H.D. Zeh, \PL {\bf A116} (1986) 9; \\
C. Kiefer, \CQG {\bf 4} (1987) 1369; ibid {\bf 6} (1989) 561.

\bibitem{dd} J.J. Halliwell, \PR {\bf D39} (1989) 2912.

\bibitem{ee} T. Padmanabhan, \PR {\bf D39} (1989) 2924.

\bibitem{ff} M. Gell-Mann and J.B. Hartle, in Complexity, Entropy and
the
Physics of Information, ed. W.H. Zurek (Addison-Wesley 1990).


\bibitem{ggg} J.P. Paz and S. Sinha, \PR {\bf D44} (1991) 1038.

\bibitem{ggga} J.P. Paz and S. Sinha, \PR  {\bf D45}
(1992) 2823.


\bibitem{hh} S. Sinha and B.L. Hu, \PR {\bf D44} (1991) 1028.

\bibitem{jj} C. Kiefer, \PR {\bf D46} (1992) 1658.

\bibitem{kk} J.J. Halliwell and S.W. Hawking, \PR {\bf D31} (1985) 1777;
\\
H.F. Dowker nad R. Laflamme, \NP {\bf B366} (1991) 209.

\bibitem{lll} O. Bertolami and J.M. Mour\~ao, \CQG {\bf 8} (1991) 1271.

\bibitem{mm} J. Hartle and S.W. Hawking, \PR {\bf D28} (1983) 2960; \\
S.W. Hawking, \NP {\bf B239} (1984) 257.

\bibitem{nn} O. Bertolami, J.M. Mour\~ao, R.F. Picken and I.P.
Volobujev, \IJMP
{\bf A6} (1991) 4149.

\bibitem{oo} P.V. Moniz and J.M. Mour\~ao, \CQG {\bf 8} (1991) 1815.

\bibitem{pp} M.C.Bento, O. Bertolami, P.V. Moniz, J.M. Mour\~ao and
P.M. S\'a,
\CQG {\bf 10} (1993) 285.

\bibitem{qq} P.V. Moniz, J.M. Mour\~ao and P.M. S\'a, \CQG {\bf 10}
(1993) 517.

\bibitem{rr} M.C.Bento and O. Bertolami, \PL {\bf B336} (1994) 6.

\bibitem{ss} O. Bertolami and P.V. Moniz, ``Decoherence of Homogeneous
and
Isotropic Geometries in the Presence of Massive Vector Fields", Lisbon
Preprint
DF/IST 12.93 (1993), in Proceedings of the III National Meeting on
Astronomy
and Astrophysics, July 1993, Lisbon, Portugal; Bulletin  Board GR-QC
9407025.


\bibitem{ss1} O. Bertolami and P.V. Moniz, ``Decoherence of Homogeneous
and
Isotropic Metrics in the Presence of Massive Vector Fields", CERN-DAMTP
report,
in: Proceedings of the Seventh Marcel Grossmann Meeting, July 1994,
Stanford University, USA; Bulletin  Board GR-QC 9409042.



\bibitem{tt} H.F. Dowker, \NP {\bf B331} (1990) 194; Ph.D. Thesis,
University
of Cambridge (unpublished).

\bibitem{uu} D. Kapetanakis, G. Koutsoumbas, A. Lukas and P. Mayr,
 Bulletin  Board HEP-TH 9403131.




\bibitem{hl} S. Habib and R. Laflamme,  \PR {\bf D42} (1990) 4056.

\bibitem{llo}  R. Laflamme and J. Louko, \PR {\bf D43} (1991) 3317.


\bibitem{cm} E. Calzetta and F. Mazzitelli,  \PR {\bf D42} (1990) 4066.


\bibitem{cm1} A. Gangui, F. Mazzitelli and E. Calzetta, \PR {\bf D43}
(1991) 1853.

\bibitem{fv} R. Feynman and F. Vernon, {\em Ann. Phys.} (N.Y.)
{\bf 24} (1963) 118.

\bibitem{sk}  J. Schwinger, {\em J. Math. Phys.} {\bf 2} (1961) 407; \\
L. Keldish, {\em Sov. Phys. JETP} {\bf 20} (1965) 1018; \\
G. Zhou, Z. Su, B. Hao and L. Yu, {\em Phys. Rep.} {\bf 118} (1985) 1;
\\
Z. Su, L. Chen, X. Yu and K. Chou, \PR {\bf B37} (1988) 9810.


\bibitem{kiefer} C. Kiefer, \CQG {\bf 8} (1991) 379.


\bibitem{ts1} T. Padmanabhan and T. Singh, \CQG {\bf 7}
(1990) 412.

\bibitem{ts2} T. Singh and T. Padmanabhan, {\em Ann. Phys. (N.Y.)}, {\bf
196}
(1989) 296.

\bibitem{new1} S. Habib, \PR {\bf D42} (1990) 2566

\bibitem{new2} A. Anderson, \PR {\bf D42} (1990) 585.

\bibitem{new3} M. Berry, {\em Philos. Trans. R. Soc.}, {\bf 287}
 (1977) 237.


\bibitem{pilati} M. Pilati, {\em Nucl. Phys.} {\bf B132} (1978) 138.



\bibitem{roman} P. Roman, `` Introduction to Quantum Field Theory'',
(Wiley,
New York, 1969).

\end{thebibliography}
\end{document}